\documentclass[aps,prx,twocolumn,superscriptaddress,longbibliography]{revtex4-1}
\usepackage{graphicx}
\usepackage[usenames]{color}
\usepackage{amsmath}
\usepackage{amssymb}

\begin{document}
\title{A continuum of compass spin models on the honeycomb lattice}

\author{Haiyuan Zou}
\affiliation{Department of Physics and Astronomy, University of
Pittsburgh, Pittsburgh, PA 15260}
\affiliation{
Department of Physics and Astronomy, George Mason University, Fairfax, VA 22030}

\author{Bo Liu}
\affiliation{Department of Physics and Astronomy, University of
Pittsburgh, Pittsburgh, PA 15260}

\author{Erhai Zhao}\email[]{ezhao2@gmu.edu}
\affiliation{
Department of Physics and Astronomy, George Mason University, Fairfax, VA 22030}

\author{W. Vincent Liu}
\affiliation{Department of Physics and Astronomy, University of
Pittsburgh, Pittsburgh, PA 15260}
\affiliation{
Wilczek Quantum Center, Zhejiang University of Technology, Hangzhou 310023, China
}
\begin{abstract}
Quantum spin models with spatially dependent interactions, known as compass models, play an important role in the study of frustrated quantum magnetism. One example is the Kitaev model on the honeycomb lattice with spin-liquid ground states and anyonic excitations. Another example is the geometrically frustrated quantum $120^\circ$ model on the same lattice whose ground state has not been unambiguously established. To generalize the Kitaev model beyond the exactly solvable limit and connect it with other compass models, we propose a new model, dubbed ``the tripod model", which contains a continuum of compass-type models. It smoothly interpolates the Ising model, the Kitaev model, and the quantum $120^\circ$ model by tuning a single parameter $\theta'$, the angle between the three legs of a tripod in the spin space. Hence it not only unifies three paradigmatic spin models, but also enables the study of their quantum phase transitions. We obtain the phase diagram of the tripod model numerically by tensor networks in the thermodynamic limit. We show that the ground state of the quantum $120^\circ$ model has long-range dimer order. Moreover, we find an extended spin-disordered (spin-liquid) phase between the dimer phase and an antiferromagnetic phase. 
The unification and solution of a continuum of frustrated spin models as outline here may be useful to exploring new domains of other quantum spin or orbital models.
\end{abstract}

\maketitle

 \section{Introduction}\label{sec:introduction}
Model Hamiltonians describing interacting spins localized on lattice sites are at the central stage in the field of quantum magnetism. A class of spin models, collectively known as the compass models~\cite{rmpcompass}, stand out owing to a unique feature they share in common: the spin exchange interactions differ for different lattice bond orientations. This is in contrast to the familiar Heisenberg model or the Ising model, where the exchange has the same form for all bonds connecting the nearest neighboring sites.  The compass models arise naturally as low energy effective Hamiltonians in Mott insulators with orbital degrees of freedom~\cite{KK1982,orbitalonly04,coldorb1,coldorb2,ZhaoLiu,Wu} as well as interacting systems with spin-orbit coupling. These highly nontrivial models are also very appealing from a pure theoretical point of view because they offer a natural arena to study frustrated quantum magnetism~\cite{quatmag1,quatmag2}. Exactly solvable compass models, the Kitaev model in particular, have played a pivotal role in stimulating the field of topological quantum computing~\cite{Kitaev20032,Kitaev20062}. The rich physics contained in compass models has been reviewed recently in Ref.~\cite{rmpcompass}.

Our work is directly motivated by two well known compass models defined on the honeycomb lattice. The first example is the Kitaev model~\cite{Kitaev20062}, where the exchange interactions between two neighboring sites are given by $\sigma^x_i\sigma^x_j$, $\sigma^y_i\sigma^y_j$, and $\sigma^z_i\sigma^z_j$ respectively. As shown by Kitaev, this model is exactly solvable and has anyonic excitations obeying fractional statistics~\cite{Kitaev20062}. The spatially dependent exchange interactions suppress long-range spin order and support a quantum spin liquid (SL) ground state, one of the most sought after exotic many-body states in condensed matter physics ~\cite{qsl1}. The Kitaev model, despite its theoretical appeal, is neither readily realized in materials nor easily simulated with synthetical quantum matter such as cold atoms on optical lattices. Recently, the hybrid Kitaev-Heisenberg model, a linear superposition of a Kitaev term and a Heisenberg term, was proposed for iridium oxides and solved numerically~\cite{Khaliullin3}. Besides the spin liquid phase, the hybrid model contains other interesting phases such as the stripe and the zigzag phase due to the competition between the two terms~\cite{Khaliullin3}. The phase diagram becomes even richer when off-diagonal spin exchange interactions are added~\cite{KeePRL,Chen1501,Khaliullin4}. 

The second example of compass models is the quantum $120^{\circ}$ model~\cite{ZhaoLiu,Wu}. It is very analogous to the Kitaev model but the spin operators $\sigma_x$, $\sigma_y$, and $\sigma_z$ for the three bond directions are replaced by three (pseudo)spin 1/2 operators $T^1$, $T^2$, and $T^3$ respectively. They form an angle of $120^\circ$ with each other on the $xz$ plane in spin space, hence the name ``the $120^{\circ}$ model." It was introduced to described the low energy physics of transition metal oxides ~\cite{Tokura21042000} with doubly degenerate orbitals, e.g. orbital-only models of $e_g$ orbitals on cubic lattice~\cite{orbitalonly04}. Later, two of us, and Wu independently, found that the $120^{\circ}$ model can be naturally realized in strongly interacting spinless $p$-orbital fermions on the honeycomb optical lattice~\cite{ZhaoLiu,Wu}. Although it is geometrically frustrated, spin wave analysis indicates that long-range order is stabilized by quantum fluctuations through the order by disorder mechanism~\cite{ZhaoLiu,Wu}. While the semiclassical spin-wave analysis is suggestive, the ground state of the $120^{\circ}$ model on honeycomb lattice remains to be identified unambiguously.

Given the apparent similarities between the Kitaev model and the $120^{\circ}$ model, it is natural to seek the conceptual and quantitative link between them. Indeed, these two models can be viewed as two instances of a more general class of compass models~\cite{test1}. In this paper, we provide a concrete construction and propose a ``super model" which contains three paradigm models, the Ising, the Kitaev and the $120^{\circ}$ model, as special limits. It only has a single tuning parameter $\theta'$ and a simple, intuitive picture for the three (pseudo)spin operators: they form a tripod in spin space as shown in Fig. 1. Analogous to tuning the tripod to raise or low a mounted camera in photography, dialing the angle $\theta'$ between the three legs takes the Ising model (tripod fully closed) smoothly to the Kitaev model (tripod open with three legs orthogonal to each other) and then to the $120^{\circ}$ model (tripod fully open with three legs in the same plane). Immediately, one conjectures that the phase diagram of this continuum compass model is highly nontrivial containing drastically different long-range ordered states as well as spin liquids.

We obtain the phase diagram of this ``super model" using tensor network states which have gained considerable success recently in the study of frustrated magnetism~\cite{wang2011spin,HaradaMERA,SSmodelTN,XiePRX14}. The results are summarized in Fig. 1 and Fig. 2. The order parameters are calculated using the tensor renormalization group (TRG) method formulated in thermodynamic limit~\cite{LevinTRG,HOTRG1}. We show that the ground state of the quantum $120^{\circ}$ model is a long-range ordered dimer phase, and a spin liquid phase exists in an extended region in our phase diagram \cite{noteSL}. The numerical results of TRG are further confirmed and crosschecked with Projected Entangled Pair States (PEPS) calculations~\cite{PEPS1,PEPS2} for finite systems, exact diagonalization, and spin wave analysis. We discuss the qualitative features of the quantum phase transitions between the spin liquid phase and the dimer phase by introducing a topological charge (spin vortex) for the dimer configuration. We further show that the proposed tripod model can in principle be simulated with Hubbard model in the Mott insulating regime, e.g., using cold atoms on optical lattice with artificial gauge fields. 

\section{The tripod model}\label{sec:model}

\begin{figure}
\includegraphics[width=0.48\textwidth]{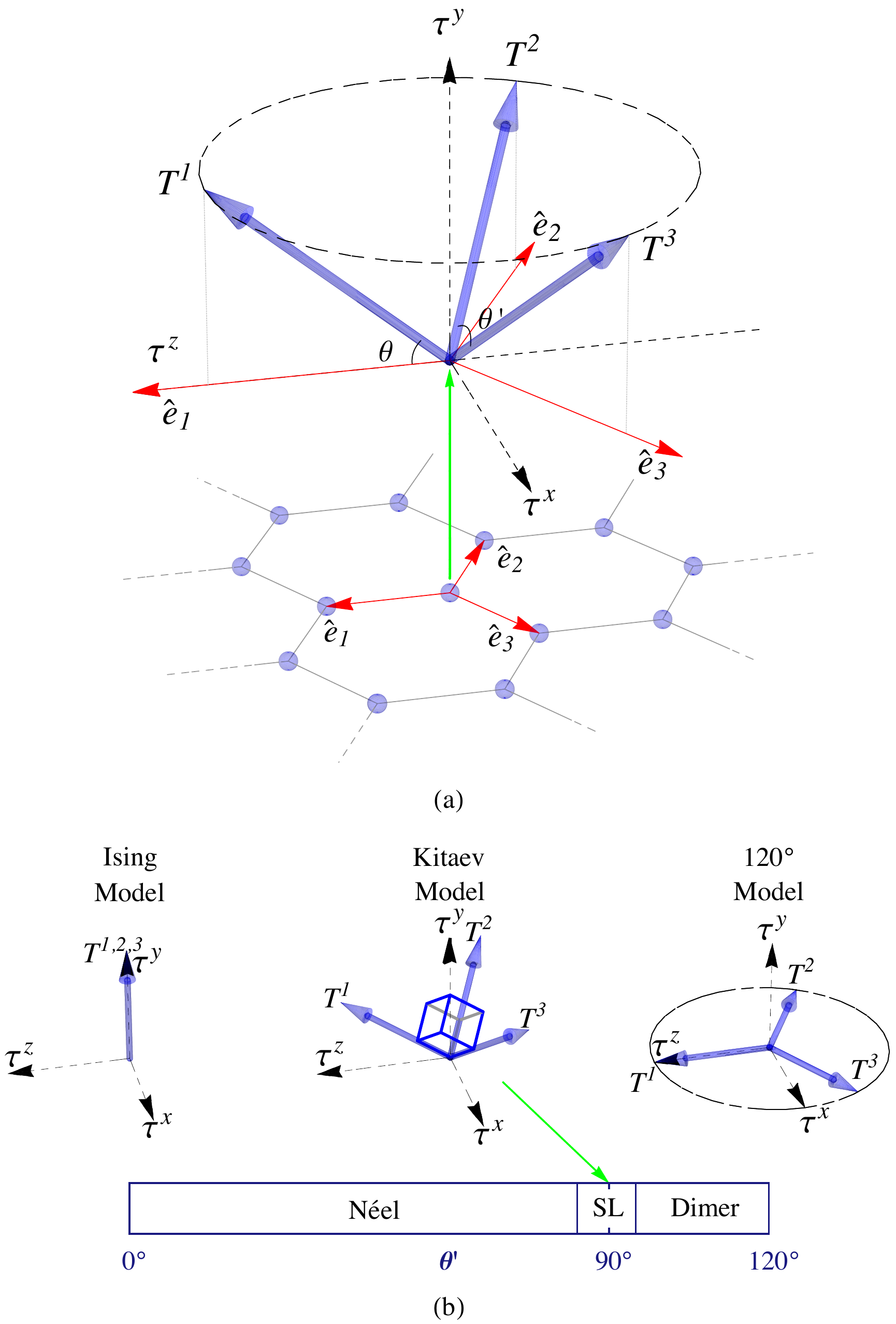}
\caption{The tripod model on the honeycomb lattice, Eq. (1). (a) The nearest neighbor spin exchange along bond direction $\mathbf{e}_\gamma$, $\gamma=1,2,3$, is defined through the spin 1/2 operator $T^\gamma $, represented by an arrow in spin space spanned by $\tau^x,\tau^y,\tau^z$. The three $T^\gamma$ can be thought as the three legs of a tripod, being tilted out of the $xz$ plane by angle $\theta$, and forming an angle $\theta'$ with each other. When projected onto the $xz$ plane, $T^\gamma$ is along the $\mathbf{e}_\gamma$ direction. (b) The schematic phase diagram of the tripod model. As the tripod is opened by increasing $\theta'$, the model starts as the Ising model at $\theta'=0$, becomes the Kitaev model at $\theta'=90^\circ$, and then the quantum $120^\circ$ model at $\theta'=120^\circ$. Three phases are identified: a N$\acute{\textrm{e}}$el ordered antiferromagnet, a spin liquid (SL), and a long-range ordered dimer phase.}
\label{spinsf}
\end{figure}

We generalize the Kitaev and the 120$^\circ$ model to the following continuum compass model defined on the two-dimensional honeycomb lattice,
\begin{equation}
H(\theta)=J\sum_{\mathbf{r},\gamma}T^\gamma_\mathbf{r}(\theta)T^\gamma_{\mathbf{r}+\mathbf{e}_\gamma}(\theta),
\label{eq:Ham}
\end{equation}
where $J>0$, and for each lattice site $\mathbf{r}$ the spin 1/2 operators are defined as
\begin{equation}
T^\gamma(\theta)=\frac{1}{2}(\tau^z\cos\phi_\gamma+\tau^x\sin\phi_\gamma)\cos\theta+\frac{1}{2}\tau^y\sin\theta
\end{equation}
with the Pauli matrices $\tau^x, \tau^y, \tau^z$. Each site $\mathbf{r}$ is coupled to its neighbors $\mathbf{r}+\mathbf{e}_\gamma$, where $\mathbf{e}_\gamma$, $\gamma=1,2,3$, denotes the three bond vectors of the honeycomb lattice. Geometrically, the three $T^\gamma$ form a tripod in the spin space as shown in Fig. 1: they are tilted from the $xz$ plane by angle $\theta$ and, when projected onto the $xz$ plane, are orientated along the corresponding bond direction $\mathbf{e}_\gamma$, i.e. at azimuthal angle $\phi_\gamma=0,2\pi/3,4\pi/3$ respectively. While $T^\gamma$ is most naturally defined through the tilting angle $\theta$, it is much more convenient to introduce another angle, $\theta'$, to discuss the various limits of $H(\theta)$. $\theta'$ is the angle between $T^1$ and $T^2$, i.e., the two adjacent legs of the tripod. And it is related to $\theta$ by trigonometry
\begin{equation}
\cos\theta'=1-\frac{3}{2}\cos^2\theta.
\end{equation}
We will take $\theta'$ as the only tuning parameter in the tripod model.

Three special limits of this model can now be identified. First of all, when
$\theta'=0$, $T^\gamma$ all collapse to $\tau_y$. The tripod is closed, and $H(\theta)$ is nothing but the Ising model,
\begin{equation}
H_I=\frac{J}{4}\sum_{\mathbf{r},\gamma} \tau^y_\mathbf{r} \tau^y_{\mathbf{r}+\mathbf{e}_\gamma}.
\end{equation}
Note that we choose $\tau_y$ as the vertical axis in spin space instead of the usual convention of $\tau_z$ so that $H(\theta)$ reduces exactly to the 120$^\circ$ model defined in our earlier work Ref.~\cite{ZhaoLiu}.  Secondly, when
$\theta'=90^{\circ}$, $H(\theta)$ reduces to the Kitaev model since the operators $T^\gamma$
are now perpendicular to each other in the spin space. We can simply identify them as $\sigma^x$, $\sigma^y$ and $\sigma^z$ (apart from a factor $1/2$) in a rotated coordinate system as illustrated in Fig. 1(b). Thirdly, for $\theta'=120^{\circ}$, $H(\theta)$ becomes the quantum $120^{\circ}$ with $T^1, T^2, T^3$ all confined within the
$xz$ plane. It can be visualized as a fully open tripod. 

As well known, the Ising model has antiferromangetic (AF) order with the order parameter 
\begin{equation}
O_1=\langle|\tau^y|\rangle/2.
\end{equation}
On the other hand side, the quantum $120^{\circ}$ model is conjectured to be long-range ordered despite the geometric frustration. We introduce the following ``order parameter" to measure the in-plane magnetization
\begin{equation}
O_2=\sqrt{\langle\tau^x\rangle^2+\langle\tau^z\rangle^2}/2.
\end{equation}

By solving $H(\theta)$ using tensor network algorithms, we compute the average spin $\langle\tau^{x,y,z}_\mathbf{r} \rangle$ in the ground state. The main results are summarized in the schematic phase diagram in Fig. 1(b). Fig. 2 shows the variation of the two order parameters introduced above as $\theta'$ is changed. The region at small $\theta'$ corresponds to the familiar N$\acute{\textrm{e}}$el order which is characteristic of the classical Ising model and illustrated in the left inset of Fig. 2. Despite the increased quantum fluctuations as $\theta'$ is increased, the N$\acute{\textrm{e}}$el ordered phase persists up to $\theta'\sim 87^\circ$. At the opposite end of large $\theta'$, we find that the long-range spin order consists of a set of ``dimers," i.e. opposite spins on neighboring sites, arranged into a periodic pattern of triangular lattice [Fig.~\ref{fig:configs2}(a)]. The triangular lattice and its enlarged unit cell becomes transparent if we introduce a topological charge [red dot in Fig.~\ref{fig:configs2}(a)] for each hexagon with spins all pointing outwards.  If we focus on one individual hexagon, e.g. the one shown in the right inset of Fig. 2, the orientations of the dimers happen to be also $60^\circ$ (or equivalently $120^\circ$) apart. We will refer to this phase simply as the ``dimer phase." In particular, it is the ground state of the quantum 120$^\circ$ model on honeycomb lattice. This point will be further discussed in Section \ref{120m}.

Sandwiched between the N$\acute{\textrm{e}}$el ordered phase and the dimer phase, a quantum spin liquid phase is stabilized for $\theta'\in [87^{\circ},94^{\circ}]$. The conclusion is mainly based on the observation from Fig. 2 that the order parameters $O_{1,2}$ in this region are nearly zero compared to those in other two phases. This conclusion is also consistent with the exactly solution of Kitaev model for $\theta'= 90^\circ$. The order parameters as functions of $\theta'$ in Fig. 2 also suggest that  the two quantum phase transitions in the tripod model may be qualitatively different. 
The gradual drop of $O_2$ at $\theta'>90^{\circ}$ indicates a continuous phase transition between the dimer phase and the spin liquid phase. In contrast, the drop of the order parameter $O_1$ at $\theta'<90^{\circ}$ is rather sharp, pointing to a likely first-order phase transition. The details of the calculations leading to these results will be discussed below in Section~\ref{sec:trg}. 

\begin{figure}
\includegraphics[width=0.47\textwidth]{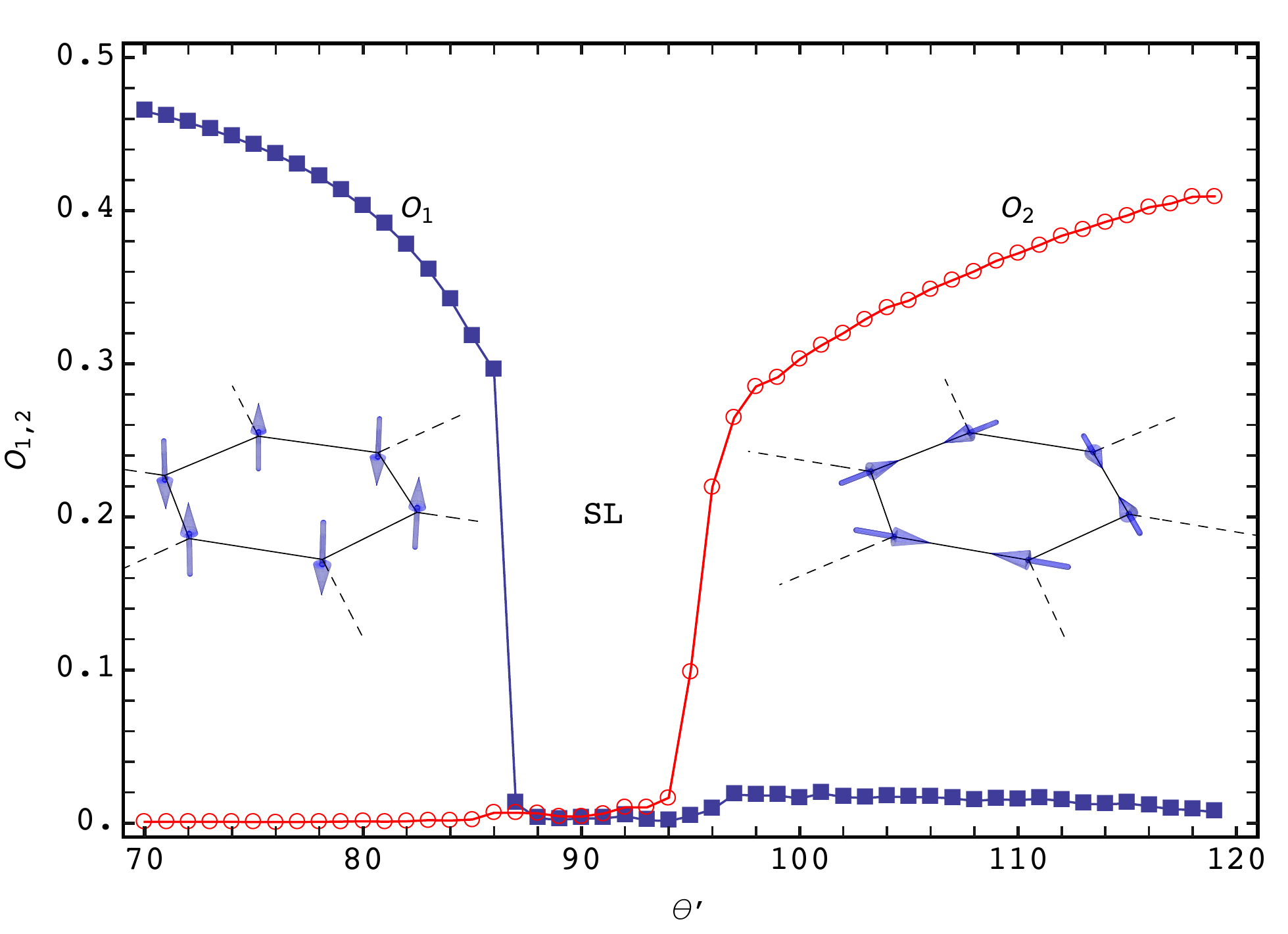}
\caption{Identifying the phases of the tripod model. The plot shows the order parameters $O_1$ (filled squares) and $O_2$ (empty circles) as a function of $\theta'$ calculated from the infinite tensor network algorithms with bond dimension $D=8$ and 6-sites unit cell. The insets illustrate the schematic spin configurations in the N$\acute{\textrm{e}}$el ordered phase (left) and the dimer phase (right) for a hexagon. For $\theta'\in [87^{\circ},94^{\circ}]$, the spin averages are zero, and the ground state is identified as a spin liquid.
}
\label{fig:info} 
\end{figure}

\section{Tensor Network algorithms}\label{sec:trg}
Recent developments of entanglement-based tensor network algorithms provide a novel, accurate approach to strongly correlated electron systems~\cite{PEPS1,PEPS2,Orus2014,iPEPS,iPEPS2}. Particularly, they have been successfully applied to frustrated quantum magnets~\cite{wang2011spin,HaradaMERA,SSmodelTN,XiePRX14} and the $t-J$ model~\cite{tJ_Troyer11,tJ_Troyer14} to yield insights previously unattainable from conventional methods. To find the phase diagram of the proposed tripod model, we employ two complementary tensor networks algorithms, one for finite-size systems and the other for infinite systems in the thermodynamic limit, to find the ground state and the order parameters. In both algorithms, the ground state wave function is constructed as a network of local tensors defined on lattice sites. Each tensor has one physical index representing the spin degree of freedom and three virtual indices, each with bond dimension $D$, describing the quantum entanglement with its three neighboring sites.

We first apply the finite PEPS algorithm~\cite{PEPS1,PEPS2,Orus2014} to solve $H(\theta)$ for a six-site system with periodic boundary conditions. The ground state energies obtained coincide with those from exact diagonalization. This suggests that PEPS is intrinsically superior compared to mean field theories when applied to frustrated spin Hamiltonians such as $H(\theta)$. The order parameters decay to zero as the ground state is approached for a finite system. Nonetheless, their decay behaviors are quite disparate for $\theta'$ values in the Ising, Kitaev, and 120$^\circ$ regions, suggesting three different phases. The details of the calculation are presented in Appendix~\ref{sec:appe1}. 

\begin{figure}
\includegraphics[width=8cm]{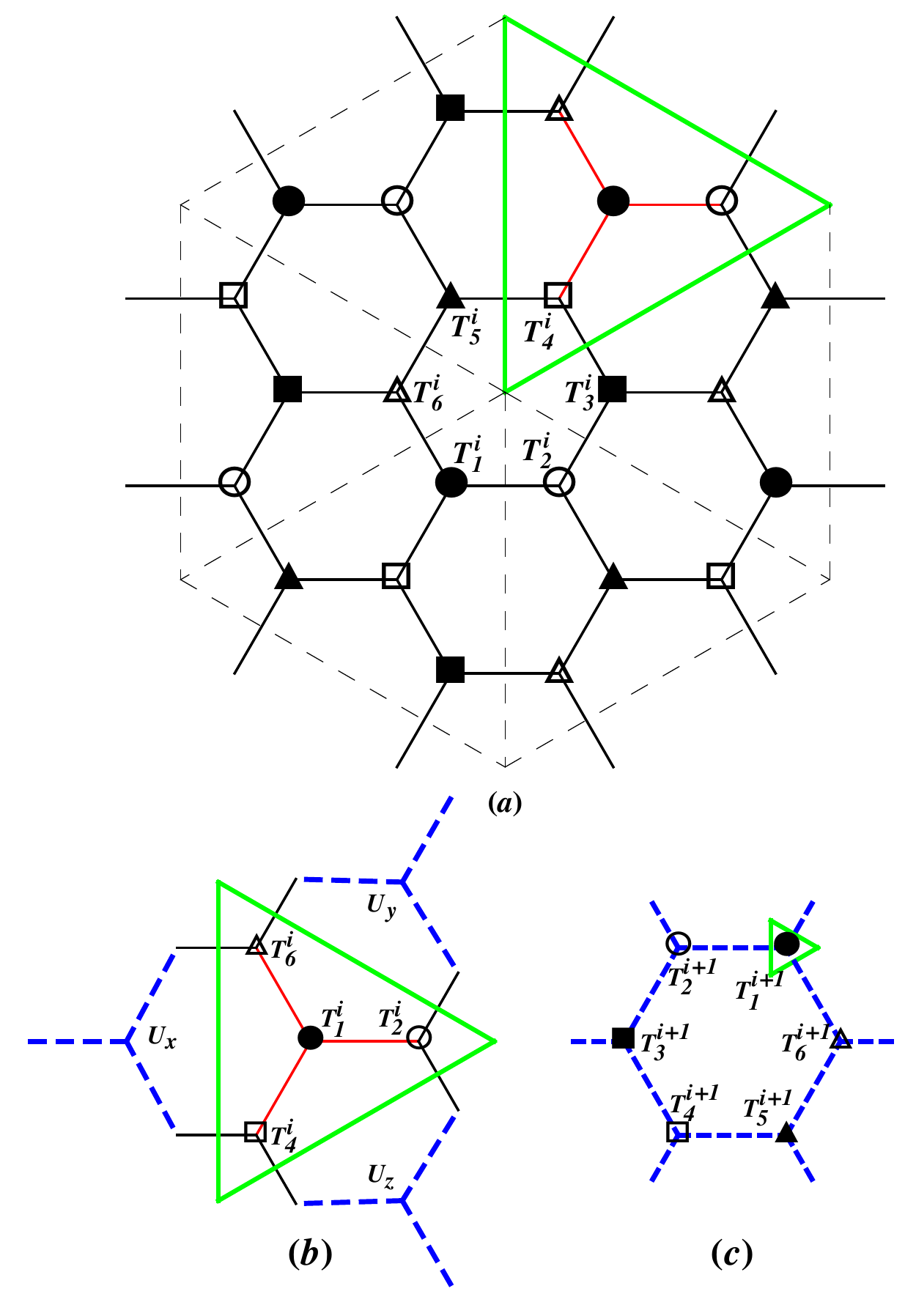}
\caption{The HOTRG procedure used to calculate the expectation value $\langle O\rangle$ after the convergence of the ground state. (a) The unit cell (the center hexagon) consists of six sites. Associated with each site is a local tensor $T^i_j$ at the $i$-th RG step.  (b) The coarse graining procedure to construct the new tensor $T^{i+1}_1$ from the old tensors $T^{i}_1$ and its neighbors $T^{i}_2, T^{i}_4, T^{i}_6$. The other five tensors are updated in the similar way. (c) The new tensors $T^{i+1}_j$ again form a honeycomb lattice. }
\label{fig:trgsp}
\end{figure}

To study the tripod model in the thermodynamic limit, we first find the converged ground state using imaginary time evolution and following the simple update scheme as described in Ref.~\cite{simpleup} which generalizes the time-evolving block decimation (TEBD)~\cite{TEBD} technique to two dimensions. For a $n$-site unit cell, e.g. a six-site unit cell shown in Fig.~\ref{fig:trgsp}(a), we need $3n/2$ different bond vectors that represent, roughly speaking, a mean-field approximation of the environment. Using these bond vectors, the simple update starts with $n$ random tensors and iterates until convergence is achieved. At the end of the calculation, the ground state $|\Psi\rangle$ is characterize by $n$ tensors $T_j$, $j\in[1,n]$. 
We then evaluate the expectation value of operator $O$, $\langle O\rangle=\langle\Psi|O|\Psi\rangle/\langle\Psi|\Psi\rangle$ which involves the (infinite) product of 
tensors $T_j$, using a real space coarse graining procedure known as higher-order tensor renormalization group (HOTRG)~\cite{HOTRG1} schematically shown in Fig.~\ref{fig:trgsp}. We outline the main steps here. At the $i$-th step of HOTRG, a local tensor, say $T^i_1$, is regrouped with its three nearest neighbor tensors ($T_2, T_4, T_6$) to form a new tensor $\tilde{T}^{i+1}_1$. More generally, for odd or even sites,
\begin{eqnarray}
\label{eq:equpdate}
\tilde{T}_o^{i+1}=\sum_\mathrm{s.l.}T_o^{i}T_2^{i}T_4^{i}T_6^{i},\\
\tilde{T}_e^{i+1}=\sum_\mathrm{s.l.}T_e^{i}T_1^{i}T_3^{i}T_5^{i},
\end{eqnarray}
where the summation is over the shared legs [abbreviated as s.l., the solid lines in Fig.~\ref{fig:trgsp}(b)] of the neighboring tensors, i.e. tensor contractions. The new tensors, each of which contains four old tensors, are of higher dimensions and truncated to have the same dimension as $T_i$ via
\begin{equation}
\label{eq:equpdate}
T_{o/e}^{i+1}=\sum_\mathrm{s.l.}\tilde{T}_{o/e}^{i+1}U_xU_yU_z,
\end{equation}
where the three projection tensors $U_{x,y,z}$, shown in Fig.~\ref{fig:trgsp}(b), are obtained as follows. Take $U_x$ as an example. First, $\tilde{T}_j$ is reshaped into matrix $\mathcal{T}_j$ with the row corresponding to the leg along the $x$-direction. Then, a matrix $\mathcal{U}$ is obtained by singular value decomposition (SVD) 
\begin{equation}
\label{eq:svdd}
\sum_{j=1}^6 \mathcal{T}_j\mathcal{T}_j^{\dagger}=\mathcal{U}\Lambda\mathcal{U}^{\dagger}.
\end{equation}
Finally, $U_x$ is obtained by truncating $\mathcal{U}$ to a given truncation dimension $\chi$ and reshaping it back to the tensor form. The new tensors $T^{i+1}$ now form exactly the same honeycomb lattice structure as the old tensors $T^i$ but represent a larger system, see Fig. ~\ref{fig:trgsp}(c). This constitutes a single RG step. By iterating the RG steps many times, the converged result of $\langle O\rangle$ well approximates the expectation value in the thermodynamic limit.

By following these procedures, we have calculated the ground state energy and the ground state expectation values of the order parameters $O_1, O_2$ for different unit cell sizes, $n=2,4,6,8$. We found that the six-site unit cell gives the lowest energy. The two-site unit cell yields results in agreement with the six-site unit cell within the parameter region $\theta'<90^{\circ}$. The four-site and eight-site unit cells, however, lead to excited states with significantly higher energy. Thus, we conclude that the six-site unit cell is the most reasonable choices for all the $\theta'$ values in the ground state calculation. In practice, one can safely use the two-site unit cell for $\theta'<90^{\circ}$ since it is significantly cheaper. The phase diagram and the spin configurations in the ordered phases shown in Fig.~\ref{fig:info} are obtained by using the two-site unit cell for $\theta'\le 90^{\circ}$ and the six-site unit cell for $\theta'>90^{\circ}$.

\section{Spin wave analysis}\label{sec:odiso}
To cross-check the TRG results, we perform the standard spin wave analysis of the tripod model. It is important to keep in mind that the validity of the spin wave theory, which can be viewed as expansion in series of $1/S$, becomes questionable in the limit of $S=1/2$. Yet the analysis offers a rough picture of the role played by geometric frustration and how the N$\acute{\textrm{e}}$el order and dimer order get destroyed by the increased quantum fluctuations. As we will show below, the estimations of the two quantum critical points from the spin wave theory turn out to be in broad agreement with the phase digram predicted by the tensor network algorithms.

The analysis starts by partitioning the honeycomb lattice into the A and B sublattice and introducing $S_\mathbf{r}=\pm T_\mathbf{r}$ for all sites on the A (B) sublattice. Then the tripod Hamiltonian acquires a suggestive form
\begin{equation}
H(\theta)=\frac{J}{2}\sum_{\mathbf{r}{\in
A},\gamma}[S^\gamma_\mathbf{r}(\theta)-S^\gamma_{\mathbf{r}+\mathbf{e}_\gamma}(\theta)]^2 + \textrm{const}.
\label{Ham2}
\end{equation}
 Here we have promoted the spin $1/2$ operator $\boldsymbol{\tau}/2$ to general spin operator $\mathbf{S}$ with spin quantum number $S$. It follows that classical ground states are massively degenerate (except for the Ising limit). Any spin configurations with $S^\gamma_\mathbf{r}(\theta)=S^\gamma_{\mathbf{r}+\mathbf{e}_\gamma}(\theta)$, i.e., the projection of $\mathbf{S}$ along the bond direction being the same for any two neighboring sites, will minimize the classical energy. This is a well known feature of compass models, see the review Ref. \cite{rmpcompass}. The special case of the classical 120$^\circ$ model on honeycomb lattice was previously discussed in Ref. \cite{Wu,nasu}. We will confine our spin wave analysis to the simple case of spatially homogeneous spin configurations ${\mathbf{S}}_\mathbf{r}=\mathbf{S}_0$ as done in Ref. \cite{nasu}. The direction of $\mathbf{S}_0$ is characterized by its polar angle $\varphi$ measured from $\tau_y$ and its  azimuthal angle $\alpha$ of $\mathbf{S}_0$ measured from $\tau_z$ in $\tau_x$-$\tau_z$ plane. 
The corresponding classical ground state energy per unit cell is \begin{equation}
\frac{E_0}{S^2 J}=-\frac{3}{2}(2\sin^2\theta\cos^2\varphi+\cos^2\theta\sin^2\varphi).
\label{eq:leading}
\end{equation}
It is interesting to note that, coincidently, at the 
Kitaev point, $\theta'=90^{\circ}$ which corresponds to $\theta=\cos^{-1}\sqrt{2/3}$, $E_0$ is completely flat and does not depend on $\varphi$. For $\theta'<90^{\circ}$, $E_0$ is minimized when $\varphi=0$ or $\pi$, corresponding to the two degenerate states with spin up or down in the N$\acute{\textrm{e}}$el ordered phase. In contrast, for $\theta'>90^{\circ}$, $E_0$ is minimized when $\varphi=\pi/2$, i.e., $\mathbf{S}_0$ lies within the $\tau_x$-$\tau_z$ plane. Therefore, the mean field theory above predicts that the tripod model has a phase transition exactly at the Kitaev point.

Applying the Holstein-Primakoff transformation~\cite{HP1} to $H(\theta)$ and expanding the resulting Hamiltonian of bosons to order $1/S$,  we compute the quantum fluctuation correction to the ground state energy for the two long-ranged ordered states respectively and find
\begin{equation}
\frac{E_1}{SJ}=\frac{1}{N}\sum_{\textbf{k},\lambda}|\omega_{\lambda}(\textbf{k})|-\Delta,
\label{eq:nlead}
\end{equation}
where $\Delta=-E_0/S^2J$, $N$ is the number of sites within the A sublattice, the $\textbf{k}$ summation is over the first Brillouin zone of the A sublattice,
and $\omega_{\lambda}(\textbf{k})$ describe the spin wave dispersion for branch $\lambda=1,2$ and they are given by the eigenvalues of the matrix
\begin{equation}
\label{eq:matrix}
\frac{1}{4}
\begin{bmatrix}
2\Delta & \beta_3^* & 0 &\beta_1^*\\
\beta_3& 2\Delta & \beta_2^* & 0\\
0 & -\beta_2 & -2\Delta & -\beta_3^*\\
-\beta_1 & 0 & -\beta_3 & -2\Delta\\
\end{bmatrix}.
\end{equation}
The expressions for $\beta_1$, $\beta_2$ and $\beta_3$ are lengthy and tabulated in Appendix~\ref{sec:appe2}. 
In what follows, we will discuss the energy $E(\varphi,\alpha)=E_0+E_1$ separately for the two distinct cases: $\theta'<90^{\circ}$ and $\theta'>90^{\circ}$.

\begin{figure}
\includegraphics[width=0.45\textwidth]{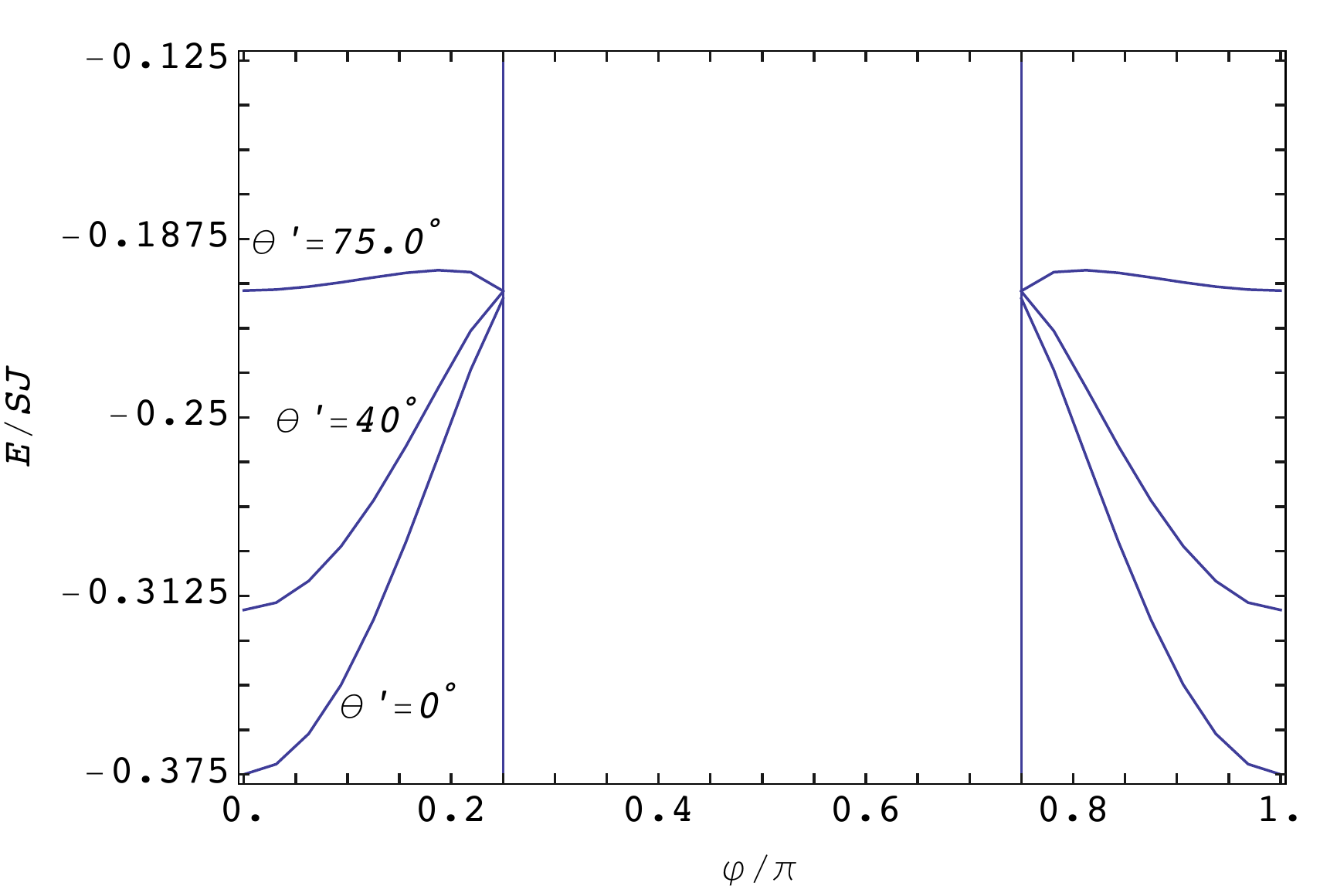}
\includegraphics[width=0.45\textwidth]{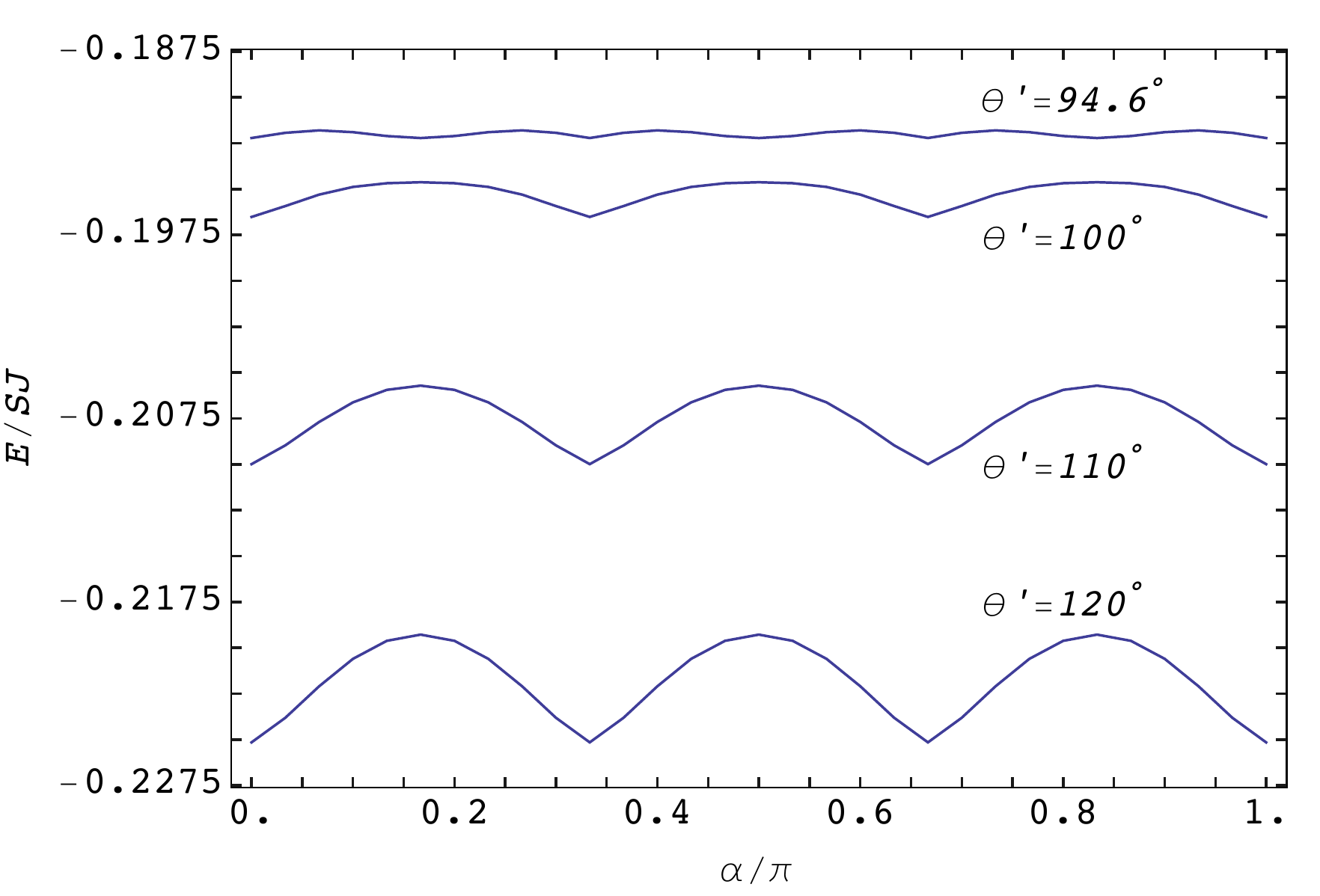}
\caption{ The energy per unit cell from leading order spin wave theory for the tripod model. 
Upper panel: $E(\phi)$ for the N$\acute{\textrm{e}}$el ordered phase. At $\theta'\sim75.0^{\circ}$, the location and number of minima of $E$ change, indicating the transition to a different phase. Lower panel: $E(\phi=\pi/2,\alpha)$ for the dimer phase with minima occurring at $\alpha_n=n\pi/{3}$. At the critical point $\theta'\sim94.6^{\circ}$, $E(\alpha)$ changes qualitatively, signals another phase transition. 
}
\label{fig:HPising120}
\end{figure}

The results for $E(\varphi)$ from the spin wave analysis are plotted in the upper panel of Fig.~\ref{fig:HPising120} for several values of $\theta'$ corresponding to the N$\acute{\textrm{e}}$el ordered phase. One notices that the fluctuations do not change qualitatively the mean field ground state. $E(\varphi)$ reaches minima still at $\varphi=0$ or $\pi$ for small $\theta'$. However, as $\theta'$ is increased, the energy $E(\varphi)$ becomes flatter. Eventually, as $\theta'=\theta'_c\sim 75.0^{\circ}$ (the top curve of Fig.~\ref{fig:HPising120}), the energies for $\varphi=\pi/4, 3\pi/4$ with proper choice of $\alpha$ become degenerate with those for $\varphi=0,\pi$. This signals the destabilization of the N$\acute{\textrm{e}}$el order by  quantum fluctuations. This occurs around $\theta'_c$,  before the Kitaev point is approached. Note that in Fig.~\ref{fig:HPising120}, only the region $\varphi\in[0,{\pi}/{4}]\cup[{3\pi}/{4},\pi]$ is shown. Outside this region (and also for $\theta'>\theta'_c$), the lowest order spin wave theory based on the Ising-like antiferromagnetic order becomes ill defined. 

For $\theta'>90^{\circ}$, the classical ground state is continuously degenerate with $\varphi=\pi/2$ but arbitrary
$\alpha\in[0,2\pi]$. Quantum fluctuations lift the degeneracy and select a long-range ordered ground state via the ``order by disorder mechanism." Such mechanism for the special case of $\theta'=120^{\circ}$, i.e. the quantum $120^{\circ}$ model, has been discussed before in Ref.~\cite{ZhaoLiu,Wu,nasu}. As shown in the lower panel of Fig.~\ref{fig:HPising120}, the same physical picture continues to hold for the tripod model for $\theta'\leq 120^{\circ}$: the energy $E$ is minimized at $\alpha_n=n\pi/{3}$ with integer $n$. However, $E(\alpha)$ becomes increasingly flat as $\theta'$ is decreased. At the critical point $\theta'=\theta'_d\sim 94.6^{\circ}$, additional minima of $E$ appear at $\alpha=\alpha_n+\pi/6$. This indicates that the long-range order gets destroyed and replaced by a new phase for $\theta'\leq \theta'_d$.

We emphasize that the simple version of spin wave theory outlined above is only intended to estimate the lower and upper critical points for the spin liquid phase, $\theta'_c$ and $\theta'_d$. It can be further improved to properly treat general classical spin configurations. We will not do it here because the large $S$ expansion by itself cannot unambiguously determine the order for our model of $S=1/2$ in the region $\theta'>90^{\circ}$.

\section{The dimer phase}\label{120m}
\begin{figure}
\includegraphics[width=7cm]{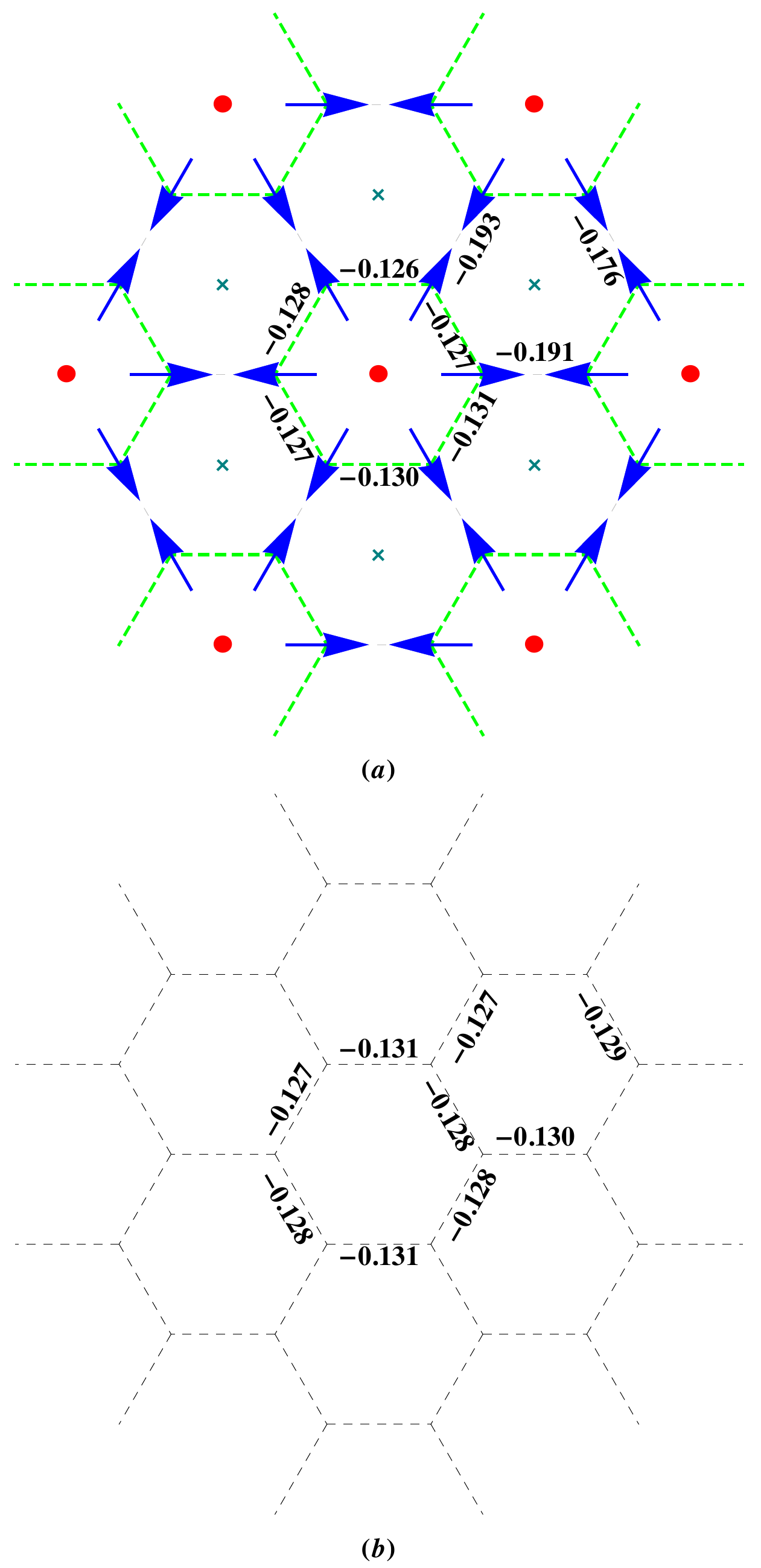}
\caption{Configurations for (a) the dimer phase ($\theta'=120^\circ$) and (b) the spin liquid phase ($\theta'=90^\circ$) of the tripod model. The number associated with each bond is $\langle T^\gamma_\mathbf{r}T^\gamma_{\mathbf{r}+\mathbf{e}_\gamma}\rangle$, i.e. the bond energy in unit of $J$, from HOTRG calculation. The arrows in (a) depict the spin direction on each site.
}
\label{fig:configs2}
\end{figure}
Previous theoretical studies of the quantum 120$^\circ$ model on the honeycomb lattice gave conflicted results. The spin wave analysis of Ref.~\cite{ZhaoLiu} assumed a homogenous ground state ${\mathbf{S}}_\mathbf{r}=\mathbf{S}_0$ and found quantum fluctuations prefer $\alpha_n=n\pi/{3}$. This led the authors to suggest that the ground state may be a simple ferromagnet of $\mathbf{S}$ with any choice of $\alpha_n$ (i.e. antiferromagnetic order in terms of the original spin $\mathbf{T}$ or $\boldsymbol{\tau}$). Ref.~\cite{Wu} considered more general (inhomogeneous) classical ground states and discovered that, within spin wave theory, the ferromagnetic state is energetically less competitive than a ``fully packed unoriented loop configuration" with the same $\alpha_n$ values. The reason is quite subtle but argued to be physically robust: the loop configuration hosts maximum number of zero modes. This result obtained from semiclassical large-$S$ expansion was conjectured to survive in the limit of $S=1/2$, i.e. the quantum 120$^\circ$ model has a ground state with the six-site plaquette order~\cite{Wu}. However, no evidence of long-range order was found in exact diagonalization (ED) studies where the spin correlation functions were computed for finite size clusters with periodic boundary conditions~\cite{nasu}. Instead, the ED results supported a trial wave function similar in spirit to the short-range resonating valence bond state, i.e., a liquid state with linear superposition of dimer covering of the lattice. Therefore, the true ground state of the quantum 120$^\circ$ model was not settled. 

Compared to these previous works, the numerical tensor network algorithm used here takes into account quantum fluctuations beyond the lowest order spin wave theory, works directly in the thermodynamic limit, and starts with unbiased (random) choice of tensors as variational parameters. It is capable of describing both the long-range ordered and the spin-disordered ground states. We find the ground state of the quantum 120$^\circ$ model is the dimer phase illustrated in Fig.~\ref{fig:configs2}(a)  where the arrows denote the direction $\hat{n}$ of spin average $\langle \boldsymbol{\tau}\rangle$ on each site, and the numbers indicate the bond energies in unit of $J$. The long-range spin order we observed agrees with the conjecture based on physical insights in Ref.~\cite{Wu}. 

We prefer the shorter, more descriptive name of ``dimer phase" adopted here because it indicates a solid (crystal) order of ``dimers," i.e. antiferromagnetically aligned spins along the bond direction, on a subset of the bonds. We propose to describe the long-range order using the vorticity or the winding number of the spin configuration around each hexagon,
\begin{equation}
\nu =\frac{1}{3} \sum_{j=1}^6 \hat{n}_j \cdot \hat{n}_{j+1},
\end{equation}
where $j$ labels the six sites of the hexagon. For example, hexagons marked by a dot in the center in Fig.~\ref{fig:configs2}(a)  correspond to $\nu=1$ where all spins on the vertices point radially outwards (corresponding to the ``loop" in Ref.~\cite{Wu}). The rest of the hexagons, each marked by a cross at the center, have $\nu=-1/2$. It then becomes apparent that the hexagons marked by dots form a triangular lattice of spin vortices. And within one unit cell of the triangular lattice, the total vorticity is zero. Note that the state shown in Fig.~\ref{fig:configs2}(a) is energetically degenerate with a state where all the spins are flipped. 

\begin{figure}
\includegraphics[width=0.48\textwidth]{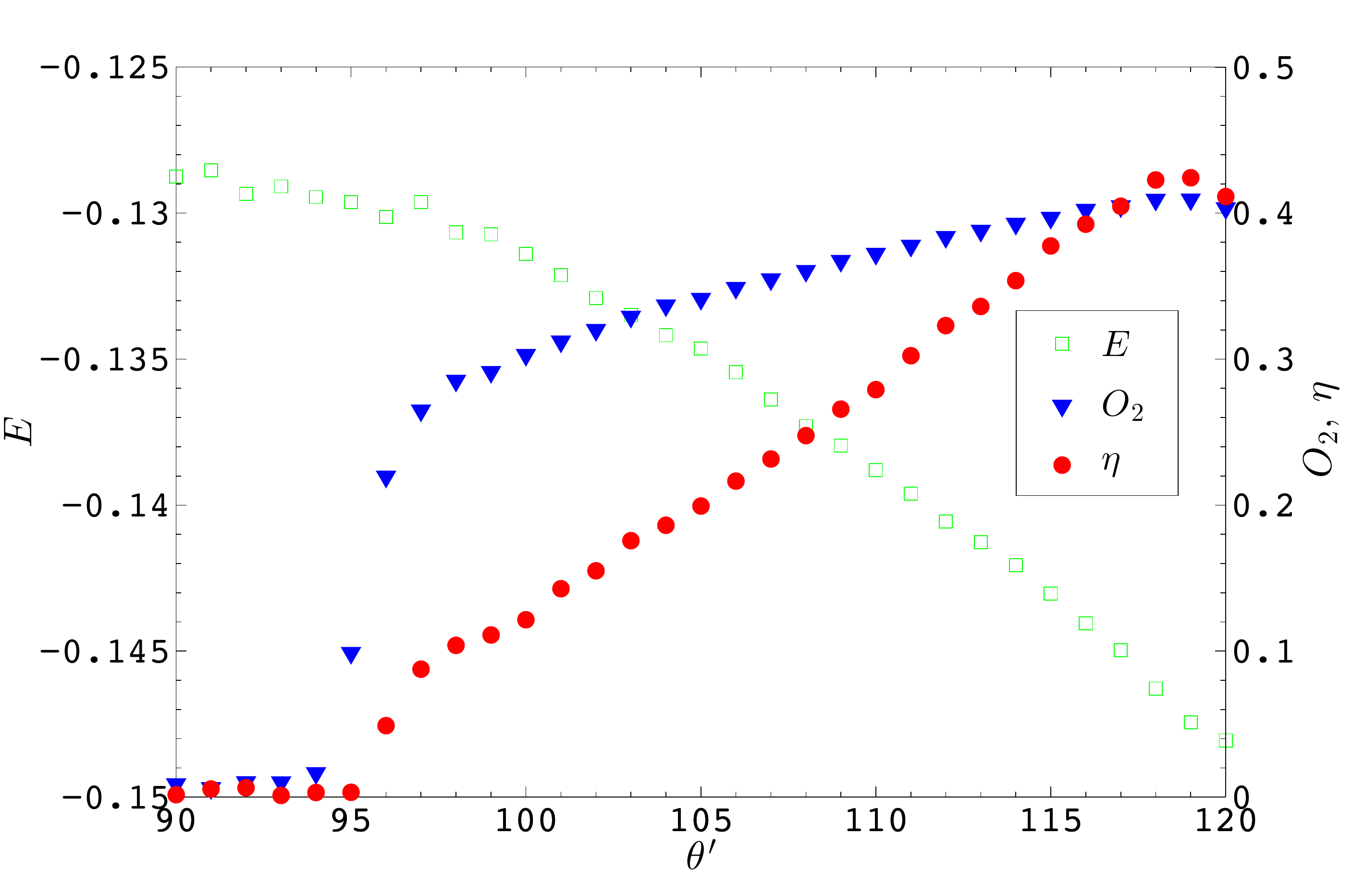}
\caption{The order parameters of the dimer phase, $O_2$ and $\eta$ (defined in the main text), and the ground state energy $E$ as functions of $\theta'$. The vanishing of $O_2$ and $\eta$ marks the transition from the dimer phase to the spin liquid phase.}
\label{fig:ztry3}
\end{figure}

By embedding the quantum 120$^\circ$ model into the more general tripod model, we are able to monitor the suppression of the dimer order and its eventual transition into the gapless spin liquid phase (phase B) of the Kitaev model. The results are summarized in Fig.~\ref{fig:ztry3}. We observe that the in-plane magnetization $O_2$ decreases continuously to zero as $\theta'$ is reduced. Meanwhile, the ground state energy $E$ steadily rises, indicating an increased degree of frustration as the Kitaev point is approached. One can measure the dimer order by introducing the energy difference $\delta E$ between the averages of two types of bonds: the ``happy" bonds (dimers) with antiparallel spins and the frustrated bonds where the two spins form an angle of $60^\circ$. Therefore, the dimensionless parameter
\begin{equation}
\eta=\delta E/E
\end{equation}
 can also serve as the order parameter for the dimer phase. As plotted in Fig.~\ref{fig:ztry3}, $\eta$ continuously drops to zero as $\theta'$ is reduced from 120$^\circ$ to $95^\circ$. Once inside the spin liquid phase, both $O_2$ and $\eta$ vanish, and the bond energies become approximately the same [see Fig.~\ref{fig:configs2}(b)]. One can view the transition from the spin liquid to
the dimer phase as condensation of spin vortices. Equivalently, when $\theta'$ is reduced, one can view the demise of the dimer order as the melting of the spin vortex lattice.
Note the bond energy shown in Fig. 5 features small fluctuations and does not strictly obey $C_6$ rotation symmetry. In our tensor network calculations, no spatially symmetry is enforced on the tensors, and the expectation values of operators are computed approximately. The fluctuations are expected to decrease as the bond dimension is increased.

\section{Potential realization}\label{sec:expe}
The tripod model can be realized from the following Hubbard
model at half filling in the Mott limit,
\begin{equation}
\label{eq:hubm}
H_{\textrm{hub}}=-\sum_{i,\gamma,\sigma\sigma'}t^\gamma_{\sigma\sigma'}f_{i,\sigma}^\dagger f_{i+e_\gamma,\sigma'}+U\sum_i n_{i,\uparrow}n_{i,\downarrow},
\end{equation}
where $f_{i,\sigma}$ annihilates a fermion with spin $\sigma$ at site $i$. The direction and spin dependent hopping matrix is related to $T^\gamma$ defined in Eq. (2) simply by
\begin{equation}
t^\gamma_{\sigma\sigma'}=t[1/2+ T^{\gamma}(\theta)]_{\sigma\sigma'}.
\end{equation}
Explicitly, they are given by
\begin{eqnarray*}
t_{\downarrow\downarrow}&=&\frac{t}{2}(1-c_1c_2),\;\;\;
t_{\uparrow\uparrow}=\frac{t}{2}(1+c_1c_2),\\
t_{\uparrow\downarrow}&=&\frac{t}{2}(c_1s_2-is_1),
t_{\downarrow\uparrow}=\frac{t}{2}(c_1s_2+is_1),
\end{eqnarray*}
where we have suppressed the superscript $\gamma$ for brevity, and
\begin{equation*}
s_1=\sin\theta, c_1=\cos\theta, s_2=\sin\phi_\gamma, c_2=\cos\phi_\gamma.
\end{equation*}
In the limit of  $U\gg t$, using second-order perturbation theory, we obtain the following effective Hamiltonian for $H_{\textrm{hub}}$
\begin{equation}
\label{eq:Hpmat}
H_\mathrm{eff}=J\sum_{\mathbf{r},\gamma}T^\gamma_\mathbf{r}(\theta)T^\gamma_{\mathbf{r}+\mathbf{e}_\gamma}(\theta)-\frac{J}{4},
\end{equation}
which is nothing but the tripod model $H(\theta)$, up to a constant term, with the superexchange $J={2t^2}/{U}$.
Note that the derivation of the effective compass Hamiltonian above does not depend on the details of the parameterization of $T^\gamma$ in terms of $\theta$ and $\phi_\gamma$. It follows that a large class of compass models, not limited to the tripod model proposed here, can be engineered on honeycomb lattice following the recipe above. 

Duan et al previously showed that the Kitaev model can be realized using cold atoms on a hexagonal optical lattice with extra laser beams~\cite{Duan}. Generalization of their idea to the case of the tripod model (and other compass models) requires spin-dependent hopping $t_{\sigma\sigma'}$ controlled by a non-Abelian gauge field or generalized spin-orbit coupling. Schemes to realize spin-orbit coupling was proposed in various approaches~\cite{WVLiu04,PZoller08,DeMarco10,PZoller11}. The realization of many have been demonstrated successfully in cold atoms experiments~\cite{Porto07,Lewenstein11,EsslingerNat14,Esslinger}. For example, spin-dependent optical lattices have been engineered using magnetic gradient modulation~\cite{Esslinger,Xupra,MGM2}. It seems possible, but challenging, to make $t_{\sigma\sigma'}$ spatially dependent. Alternatively, the tripod model proposed here may be emulated using other artificial quantum systems such as superconducting quantum circuits~\cite{SCcircuit}.

\section{Outlook}\label{sec:disc}

The tripod model introduced in this paper encompasses three well known models of quantum magnetism: the Ising model, the Kitaev model and the $120^\circ$ model. We established its (zero temperature) phase diagram using tensor network algorithms. This amounts to solving a continuum of frustrated spin models with spatially dependent exchange interactions. In particular, we found an extended spin liquid phase around the Kitaev point, and a dimer phase for large values of angle $\theta'$ including the quantum $120^\circ$ model. The two quantum critical points obtained from tensor network states agree roughly with estimations from spin wave theory. 

Our work only scratches the surface of the rich physics contained in the tripod model. Here we mention just a few open questions to be addressed in future work. First of all, it is desirable to develop a field-theoretical description of the continuous phase transition between the spin liquid phase and the long-range ordered dimer phase, based on the intuitive picture of spin vortices introduced in Section V. Secondly, the tripod model, like other compass models, has very interesting emergent symmetry properties including intermediate symmetries midway between the global and local symmetries \cite{rmpcompass}. Consequently, the excitation spectrum is expected to contain zero modes and/or flat bands. It is therefore valuable to understand the excitation spectra of the long-range order phases by going beyond the ground state analysis here. Thirdly, the finite temperature properties of the tripod model deserve a separate study. The classical limit of the tripod model is known to be highly nontrivial. The effects of thermal fluctuations and the ``order by disorder" mechanism have been investigated in Ref.~\cite{nasu} for the classical $120^\circ$ model.  Finally, we have only focused on the case of $J_\gamma=J$ here. From the Kitaev model, we know that a gapped spin liquid phase (phase A) takes over when the asymmetry in $J_\gamma$ grows large. Thus one expects that further generalization of the tripod model to general values of $J_\gamma$ may uncover new interesting phases.
 
To conclude, we hope our results can stimulate further application of tensor network algorithms to frustrated spin models as well as spin-orbital models describing transition metal oxides. We also hope our introduction of the tripod model can inspire alternative proposals to extend the Kitaev model or realize compass models in artificial quantum systems such as cold atoms on optical lattices or superconducting circuits.

\section*{Acknowledgments}
We thank Jiyao Chen, Arun Paramekanti, Zhiyuan Xie, and Congjun Wu for helpful discussions. This work is supported by AFOSR Grant No. FA9550-16-1-0006 (H.Z., B.L., E.Z., and W.V.L.), NSF Grant No. PHY-1205504 (H.Z. and E.Z.), and jointly by U.S. ARO Grant No. W911NF-11-1-0230, the Pittsburgh Foundation and its Charles E. Kaufman Foundation, and the Overseas Scholar Collaborative Program of NSF of China No. 11429402 sponsored by Peking University (W.V.L.). W.V.L is grateful for the hospitality of KITP UCSB where part of this manuscript was written with support from NSF PHY11-25915. Publication of this article is funded by the George Mason University Libraries Open Access Publishing Fund, and the University Library System, University of Pittsburgh.

\begin{appendix}
\section{Tensor network algorithms}
\label{sec:appe1}
\begin{figure}
\includegraphics[width=0.49\textwidth]{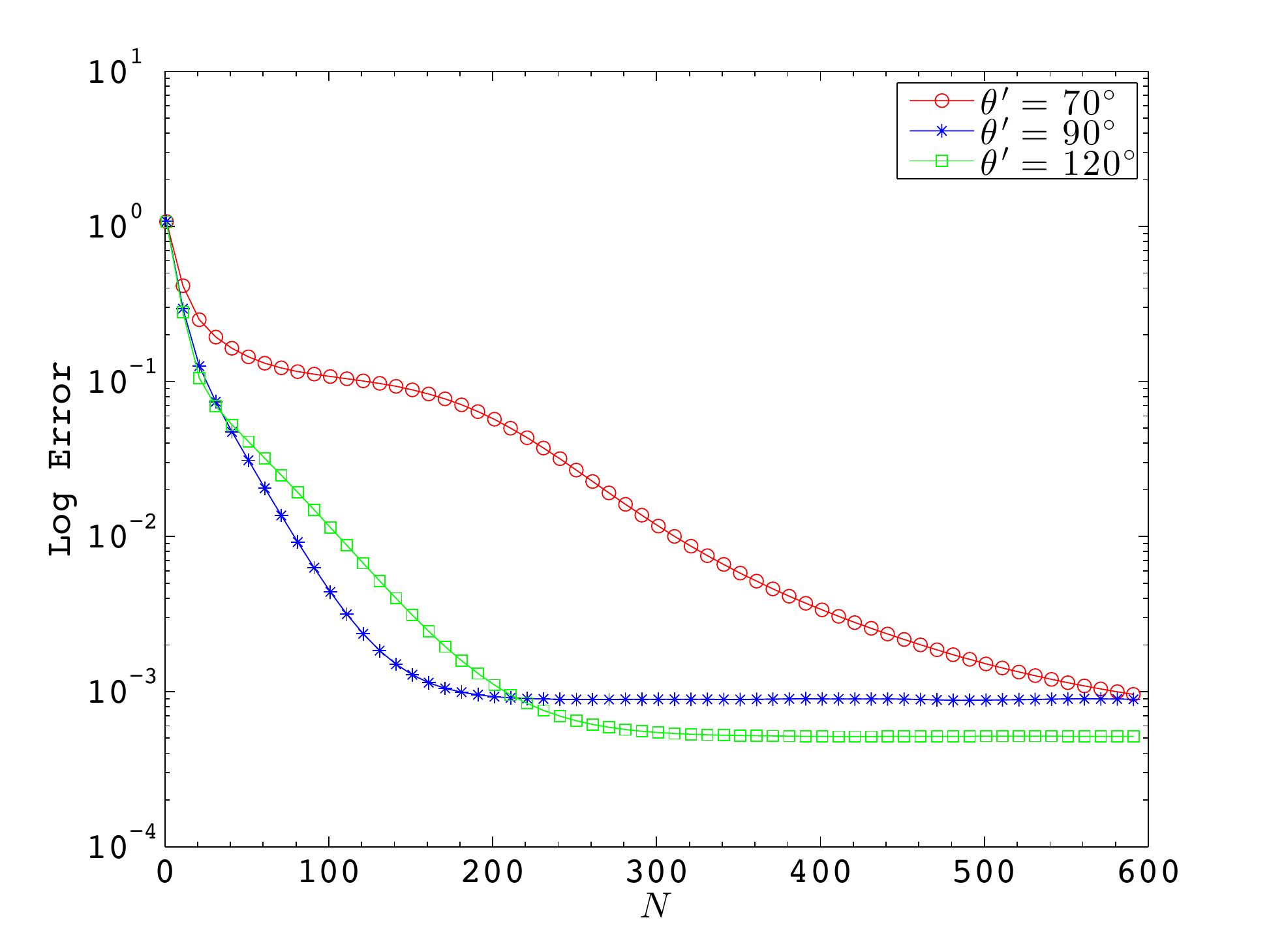}\\
\includegraphics[width=0.49\textwidth]{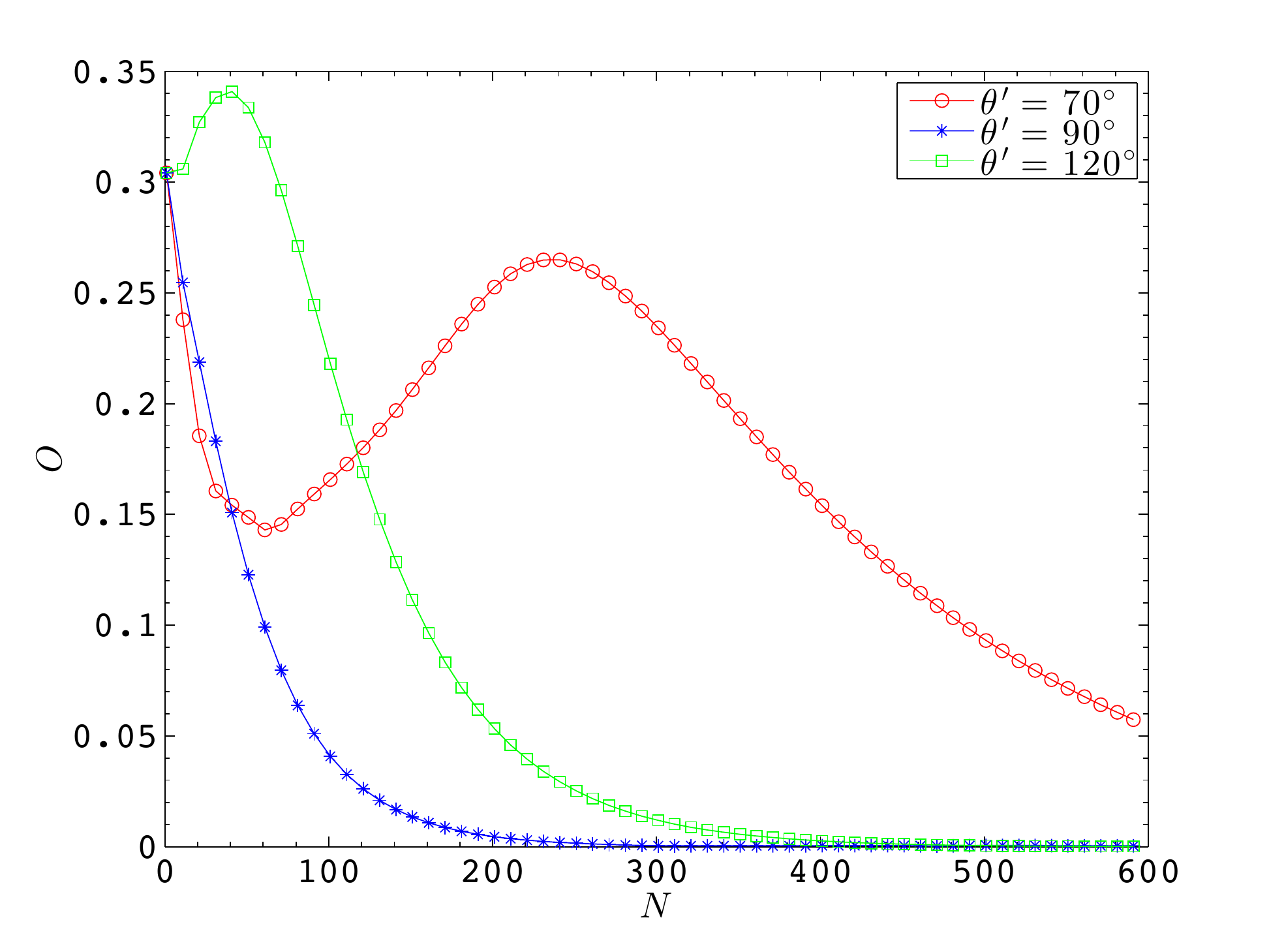}
\caption{Finite PEPS results for a six-site cluster showing the errors of ground state energy (relative to that from the exact diagonalization) and the order parameter $O$ for three different values of $\theta'$. }
\label{PEPS}
\end{figure}

The finite PEPS algorithm is a powerful numerical approach for two-dimensional quantum spin systems \cite{PEPS1,PEPS2,Orus2014}. 
For the tripod model, we construct the usual PEPS wave function starting from six random rank-four tensors with virtual bond dimension $D=3$. The tensors are then optimized through recursive imaginary-time evolution with time step $\tau=0.01$ on all the links. Once the wave function is converged, we calculate the ground state energy and the expectation value of the combined order parameter $O=\sqrt{O_1^2+O_2^2}$. The results are shown in Fig.~\ref{PEPS} for three typical values of $\theta'$ (corresponding to the three different phases found in the thermodynamic limit). For the small system size considered here (six sites with periodic boundary conditions), $O$ vanishes as the wave function converges to the ground state. Nonetheless, a noticeable peak of $O$  during the time evolution can serve as the indication for spin order. For all the three cases, the energies converge to the exact diagonalization result up to a relative error of $10^{-3}$  (the upper panel of Fig.~\ref{PEPS}). The peaks of $O$ in the lower panel of Fig.~\ref{PEPS} for the cases $\theta'=70^{\circ}$ and $\theta'=120^{\circ}$ point to the N$\acute{\textrm{e}}$el order phase and the dimer phase respectively, while the monotonic decay of $O$ for the case $\theta'=90^{\circ}$ suggests a spin liquid state.

Within the many variants of tensor network algorithms, a typical way to find the phase diagram of quantum spin models is the infinite PEPS (iPEPS) method~\cite{iPEPS,iPEPS2}. The iPEPS \textit{ansatz} on the honeycomb lattice usually proceeds by mapping the lattice to a square lattice and evaluating the effective environment by contraction schemes such as infinite matrix product states~\cite{iPEPS} or corner transfer matrices~\cite{CTM1,CTM2}. For instance, the phases of Kitaev-Heisenberg model~\cite{iPEPSKH} and the SU(4) symmetric Kugel-Khomskii model~\cite{KKprx} have been studied via the iPEPS ansatz with a $2\times 2$ or $4\times 4$ unit cell. However, the contraction scheme for a six-site (hexagonal) unit cell on the honeycomb lattice is tedious and expensive, especially for the corner transfer matrices scheme. 

\begin{figure}
\includegraphics[width=0.48\textwidth]{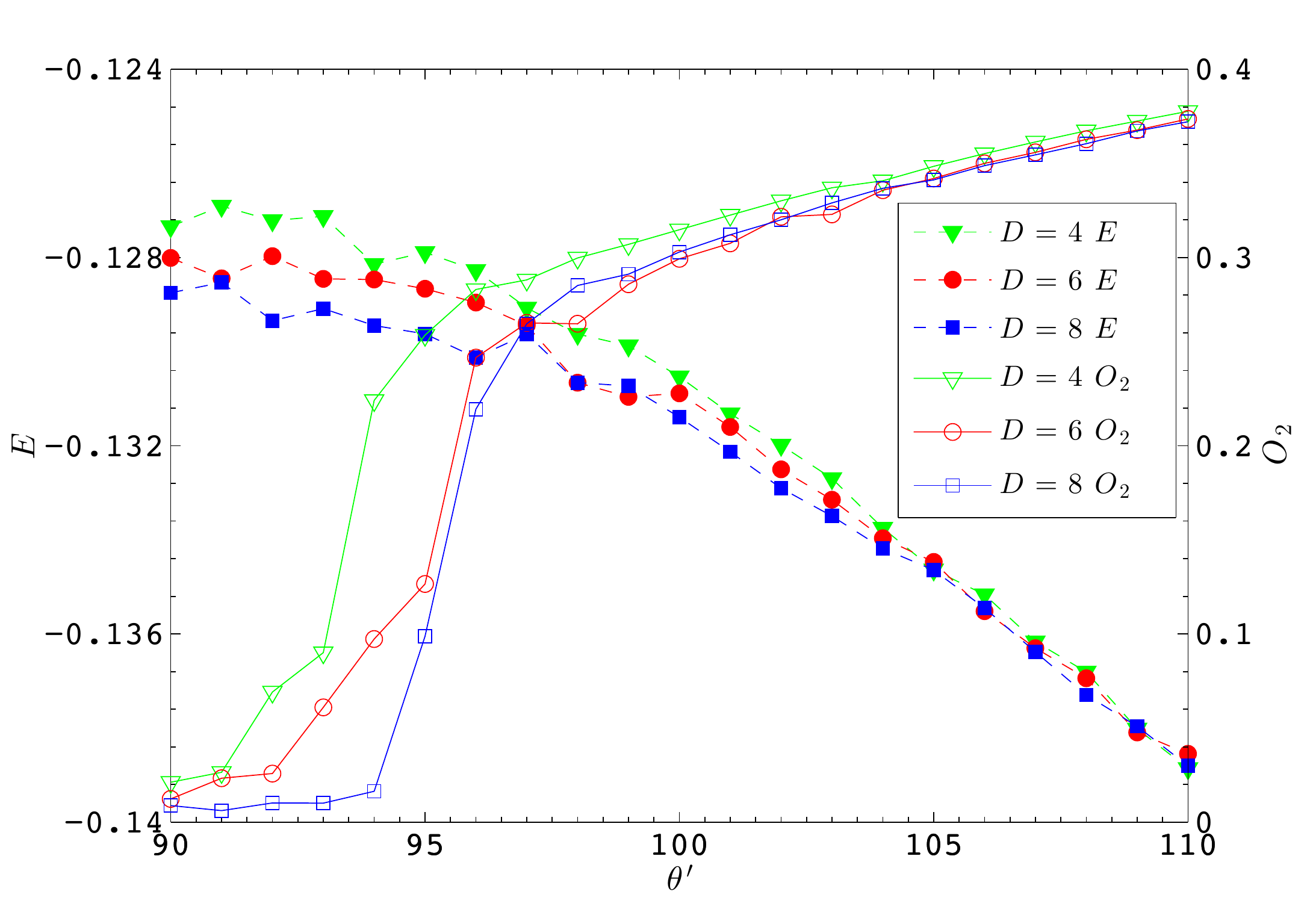}\\
\caption{The ground state energy $E$ and the order parameter $O_2$ computed from HOTRG at virtual bond dimension $D=4,6,8$ and truncation dimension $\chi=8$.}
\label{fig:Dbond}
\end{figure}

For this reason, we adopt the simple tensor update scheme and evaluate the contraction using the higher-order tensor renormalization group (HOTRG) method as explained in the main text.  The simple update~\cite{simpleup} generalizes the time-evolving block decimation~\cite{TEBD} technique to two dimensional quantum systems by introducing the bond vectors to represent the mean-field environment for local tensors. We set the imaginary time step $\tau=0.01$ and the number of iterations is generally around $10^5$ (smaller time step does not improve the numerical result significantly). The accuracy of the HOTRG method is controlled by the virtual bond dimension $D$. By systematically increasing $D$, the quantum entanglement between neighboring sites is better taken into account, yielding a more accurate ground state. For example, Fig.~\ref{fig:Dbond} shows that the order parameter $O_2$ vanishes when $D$ is increased to 8 in the region $\theta'<94^{\circ}$, suggesting a spin liquid ground state. One notices that the variations of the ground state energy with $\theta'$ within the spin liquid phase is larger than those in the long-range ordered phase, especially for smaller $D$ values. This is due to the strong quantum fluctuations intrinsic to the spin liquid. Cross-checking the HOTRG calculations here to those using Second Renormalization Group~\cite{SRG} which takes into account the entanglement between the system and the environment deserves a future study.

\section{Spin wave theory}
\label{sec:appe2}

For the two long-range ordered phases, the $1/S$ corrections to the energy are obtained by diagonalizing the matrix Eq.~\eqref{eq:matrix} with different form factors $\beta_1$, $\beta_2$, and $\beta_3$. For the N$\acute{\textrm{e}}$el ordered phase, they are given by
 \begin{eqnarray*}
 \beta_1&=&\sum_{j=1}^3(b_j+ia_j)^2e^{i\textbf{k}\cdot\hat{e}_j},\\
  \beta_2&=&\sum_{j=1}^3(b_j+ia_j)^2e^{-i\textbf{k}\cdot\hat{e}_j},\\
  \beta_3&=&-\sum_{j=1}^3(b_j^2+a_j^2)e^{i\textbf{k}\cdot\hat{e}_j}.\\
 \end{eqnarray*}
And for the dimer phase,
 \begin{eqnarray*}
 \beta_1&=&\sum_{j=1}^3(a_j+ib_j)^2e^{i\textbf{k}\cdot\hat{e}_j},\\
  \beta_2&=&\sum_{j=1}^3(a_j+ib_j)^2e^{-i\textbf{k}\cdot\hat{e}_j},\\
  \beta_3&=&-\sum_{j=1}^3(a_j^2+b_j^2)e^{i\textbf{k}\cdot\hat{e}_j}.\\
 \end{eqnarray*}
Here, $a_j$ and $b_j$ are related to the parameter $\theta$, $\varphi$, and $\alpha$ defined in the main text through 
\begin{eqnarray*}
 a_1&=&\cos\theta\cos\varphi\cos\alpha-\sin\theta\sin\varphi,\\
  a_2&=&\cos\theta\cos\varphi\cos(\alpha-\frac{2\pi}{3})-\sin\theta\sin\varphi,\\
  a_3&=&\cos\theta\cos\varphi\cos(\alpha-\frac{4\pi}{3})-\sin\theta\sin\varphi,\\
 b_1&=&-\cos\theta\sin\alpha,\\
  b_2&=&-\cos\theta\sin(\alpha-\frac{2\pi}{3}),\\
  b_3&=&-\cos\theta\sin(\alpha-\frac{4\pi}{3}).\\
  \end{eqnarray*}

\end{appendix}

\bibliography{compass120_revised.bbl}

\begin{thebibliography}{53}%
\makeatletter
\providecommand \@ifxundefined [1]{%
 \@ifx{#1\undefined}
}%
\providecommand \@ifnum [1]{%
 \ifnum #1\expandafter \@firstoftwo
 \else \expandafter \@secondoftwo
 \fi
}%
\providecommand \@ifx [1]{%
 \ifx #1\expandafter \@firstoftwo
 \else \expandafter \@secondoftwo
 \fi
}%
\providecommand \natexlab [1]{#1}%
\providecommand \enquote  [1]{``#1''}%
\providecommand \bibnamefont  [1]{#1}%
\providecommand \bibfnamefont [1]{#1}%
\providecommand \citenamefont [1]{#1}%
\providecommand \href@noop [0]{\@secondoftwo}%
\providecommand \href [0]{\begingroup \@sanitize@url \@href}%
\providecommand \@href[1]{\@@startlink{#1}\@@href}%
\providecommand \@@href[1]{\endgroup#1\@@endlink}%
\providecommand \@sanitize@url [0]{\catcode `\\12\catcode `\$12\catcode
  `\&12\catcode `\#12\catcode `\^12\catcode `\_12\catcode `\%12\relax}%
\providecommand \@@startlink[1]{}%
\providecommand \@@endlink[0]{}%
\providecommand \url  [0]{\begingroup\@sanitize@url \@url }%
\providecommand \@url [1]{\endgroup\@href {#1}{\urlprefix }}%
\providecommand \urlprefix  [0]{URL }%
\providecommand \Eprint [0]{\href }%
\providecommand \doibase [0]{http://dx.doi.org/}%
\providecommand \selectlanguage [0]{\@gobble}%
\providecommand \bibinfo  [0]{\@secondoftwo}%
\providecommand \bibfield  [0]{\@secondoftwo}%
\providecommand \translation [1]{[#1]}%
\providecommand \BibitemOpen [0]{}%
\providecommand \bibitemStop [0]{}%
\providecommand \bibitemNoStop [0]{.\EOS\space}%
\providecommand \EOS [0]{\spacefactor3000\relax}%
\providecommand \BibitemShut  [1]{\csname bibitem#1\endcsname}%
\let\auto@bib@innerbib\@empty
\bibitem [{\citenamefont {Nussinov}\ and\ \citenamefont {van~den
  Brink}(2015)}]{rmpcompass}%
  \BibitemOpen
  \bibfield  {author} {\bibinfo {author} {\bibfnamefont {Zohar}\ \bibnamefont
  {Nussinov}}\ and\ \bibinfo {author} {\bibfnamefont {Jeroen}\ \bibnamefont
  {van~den Brink}},\ }\bibfield  {title} {\enquote {\bibinfo {title}
  {\textit{Compass Models: Theory and Physical Motivations}},}\ }\href
  {\doibase 10.1103/RevModPhys.87.1} {\bibfield  {journal} {\bibinfo  {journal}
  {Rev. Mod. Phys.}\ }\textbf {\bibinfo {volume} {87}},\ \bibinfo {pages} {1}
  (\bibinfo {year} {2015})}\BibitemShut {NoStop}%
\bibitem [{\citenamefont {Kugel'}\ and\ \citenamefont
  {Khomskii}(1982)}]{KK1982}%
  \BibitemOpen
  \bibfield  {author} {\bibinfo {author} {\bibfnamefont {Kliment~I}\
  \bibnamefont {Kugel'}}\ and\ \bibinfo {author} {\bibfnamefont {D~I}\
  \bibnamefont {Khomskii}},\ }\bibfield  {title} {\enquote {\bibinfo {title}
  {\textit{The Jahn-Teller Effect and Magnetism: Transition Metal
  Compounds}},}\ }\href {http://stacks.iop.org/0038-5670/25/i=4/a=R03}
  {\bibfield  {journal} {\bibinfo  {journal} {Soviet Physics Uspekhi}\ }\textbf
  {\bibinfo {volume} {25}},\ \bibinfo {pages} {231} (\bibinfo {year}
  {1982})}\BibitemShut {NoStop}%
\bibitem [{\citenamefont {van~den Brink}(2004)}]{orbitalonly04}%
  \BibitemOpen
  \bibfield  {author} {\bibinfo {author} {\bibfnamefont {Jeroen}\ \bibnamefont
  {van~den Brink}},\ }\bibfield  {title} {\enquote {\bibinfo {title}
  {\textit{Orbital-Only Models: Ordering and Excitations}},}\ }\href
  {http://stacks.iop.org/1367-2630/6/i=1/a=201} {\bibfield  {journal} {\bibinfo
   {journal} {New Journal of Physics}\ }\textbf {\bibinfo {volume} {6}},\
  \bibinfo {pages} {201} (\bibinfo {year} {2004})}\BibitemShut {NoStop}%
\bibitem [{\citenamefont {Liu}\ and\ \citenamefont {Wu}(2006)}]{coldorb1}%
  \BibitemOpen
  \bibfield  {author} {\bibinfo {author} {\bibfnamefont {W.~Vincent}\
  \bibnamefont {Liu}}\ and\ \bibinfo {author} {\bibfnamefont {Congjun}\
  \bibnamefont {Wu}},\ }\bibfield  {title} {\enquote {\bibinfo {title}
  {\textit{Atomic Matter of Nonzero-Momentum Bose-Einstein Condensation and
  Orbital Current Order}},}\ }\href {\doibase 10.1103/PhysRevA.74.013607}
  {\bibfield  {journal} {\bibinfo  {journal} {Phys. Rev. A}\ }\textbf {\bibinfo
  {volume} {74}},\ \bibinfo {pages} {013607} (\bibinfo {year}
  {2006})}\BibitemShut {NoStop}%
\bibitem [{\citenamefont {Wu}\ and\ \citenamefont
  {Das~Sarma}(2008)}]{coldorb2}%
  \BibitemOpen
  \bibfield  {author} {\bibinfo {author} {\bibfnamefont {Congjun}\ \bibnamefont
  {Wu}}\ and\ \bibinfo {author} {\bibfnamefont {S.}~\bibnamefont {Das~Sarma}},\
  }\bibfield  {title} {\enquote {\bibinfo {title} {\textit{${p}_{x,y}$-Orbital
  Counterpart of Graphene: Cold Atoms in the Honeycomb Optical Lattice}},}\
  }\href {\doibase 10.1103/PhysRevB.77.235107} {\bibfield  {journal} {\bibinfo
  {journal} {Phys. Rev. B}\ }\textbf {\bibinfo {volume} {77}},\ \bibinfo
  {pages} {235107} (\bibinfo {year} {2008})}\BibitemShut {NoStop}%
\bibitem [{\citenamefont {Zhao}\ and\ \citenamefont {Liu}(2008)}]{ZhaoLiu}%
  \BibitemOpen
  \bibfield  {author} {\bibinfo {author} {\bibfnamefont {Erhai}\ \bibnamefont
  {Zhao}}\ and\ \bibinfo {author} {\bibfnamefont {W.~Vincent}\ \bibnamefont
  {Liu}},\ }\bibfield  {title} {\enquote {\bibinfo {title} {\textit{Orbital
  Order in Mott Insulators of Spinless $p$-Band Fermions}},}\ }\href {\doibase
  10.1103/PhysRevLett.100.160403} {\bibfield  {journal} {\bibinfo  {journal}
  {Phys. Rev. Lett.}\ }\textbf {\bibinfo {volume} {100}},\ \bibinfo {pages}
  {160403} (\bibinfo {year} {2008})}\BibitemShut {NoStop}%
\bibitem [{\citenamefont {Wu}(2008)}]{Wu}%
  \BibitemOpen
  \bibfield  {author} {\bibinfo {author} {\bibfnamefont {Congjun}\ \bibnamefont
  {Wu}},\ }\bibfield  {title} {\enquote {\bibinfo {title} {\textit{Orbital
  Ordering and Frustration of $p$-Band Mott Insulators}},}\ }\href {\doibase
  10.1103/PhysRevLett.100.200406} {\bibfield  {journal} {\bibinfo  {journal}
  {Phys. Rev. Lett.}\ }\textbf {\bibinfo {volume} {100}},\ \bibinfo {pages}
  {200406} (\bibinfo {year} {2008})}\BibitemShut {NoStop}%
\bibitem [{\citenamefont {Ferrero}\ \emph {et~al.}(2003)\citenamefont
  {Ferrero}, \citenamefont {Becca},\ and\ \citenamefont {Mila}}]{quatmag1}%
  \BibitemOpen
  \bibfield  {author} {\bibinfo {author} {\bibfnamefont {Michel}\ \bibnamefont
  {Ferrero}}, \bibinfo {author} {\bibfnamefont {Federico}\ \bibnamefont
  {Becca}}, \ and\ \bibinfo {author} {\bibfnamefont {Fr\'ed\'eric}\
  \bibnamefont {Mila}},\ }\bibfield  {title} {\enquote {\bibinfo {title}
  {\textit{Freezing and Large Time Scales Induced by Geometrical
  Frustration}},}\ }\href {\doibase 10.1103/PhysRevB.68.214431} {\bibfield
  {journal} {\bibinfo  {journal} {Phys. Rev. B}\ }\textbf {\bibinfo {volume}
  {68}},\ \bibinfo {pages} {214431} (\bibinfo {year} {2003})}\BibitemShut
  {NoStop}%
\bibitem [{\citenamefont {Capponi}\ \emph {et~al.}(2004)\citenamefont
  {Capponi}, \citenamefont {L\"auchli},\ and\ \citenamefont
  {Mambrini}}]{quatmag2}%
  \BibitemOpen
  \bibfield  {author} {\bibinfo {author} {\bibfnamefont {Sylvain}\ \bibnamefont
  {Capponi}}, \bibinfo {author} {\bibfnamefont {Andreas}\ \bibnamefont
  {L\"auchli}}, \ and\ \bibinfo {author} {\bibfnamefont {Matthieu}\
  \bibnamefont {Mambrini}},\ }\bibfield  {title} {\enquote {\bibinfo {title}
  {\textit{Numerical Contractor Renormalization Method for Quantum Spin
  Models}},}\ }\href {\doibase 10.1103/PhysRevB.70.104424} {\bibfield
  {journal} {\bibinfo  {journal} {Phys. Rev. B}\ }\textbf {\bibinfo {volume}
  {70}},\ \bibinfo {pages} {104424} (\bibinfo {year} {2004})}\BibitemShut
  {NoStop}%
\bibitem [{\citenamefont {Kitaev}(2003)}]{Kitaev20032}%
  \BibitemOpen
  \bibfield  {author} {\bibinfo {author} {\bibfnamefont {A.Yu.}\ \bibnamefont
  {Kitaev}},\ }\bibfield  {title} {\enquote {\bibinfo {title}
  {\textit{Fault-Tolerant Quantum Computation by Anyons}},}\ }\href@noop {}
  {\bibfield  {journal} {\bibinfo  {journal} {Annals of Physics}\ }\textbf
  {\bibinfo {volume} {303}},\ \bibinfo {pages} {2} (\bibinfo {year}
  {2003})}\BibitemShut {NoStop}%
\bibitem [{\citenamefont {Kitaev}(2006)}]{Kitaev20062}%
  \BibitemOpen
  \bibfield  {author} {\bibinfo {author} {\bibfnamefont {Alexei}\ \bibnamefont
  {Kitaev}},\ }\bibfield  {title} {\enquote {\bibinfo {title} {\textit{Anyons
  in an Exactly Solved Model and Beyond}},}\ }\href@noop {} {\bibfield
  {journal} {\bibinfo  {journal} {Annals of Physics}\ }\textbf {\bibinfo
  {volume} {321}},\ \bibinfo {pages} {2} (\bibinfo {year} {2006})}\BibitemShut
  {NoStop}%
\bibitem [{\citenamefont {Balents}(2010)}]{qsl1}%
  \BibitemOpen
  \bibfield  {author} {\bibinfo {author} {\bibfnamefont {Leon}\ \bibnamefont
  {Balents}},\ }\bibfield  {title} {\enquote {\bibinfo {title} {\textit{Spin
  Liquids in Frustrated Magnets}},}\ }\href
  {http://dx.doi.org/10.1038/nature08917} {\bibfield  {journal} {\bibinfo
  {journal} {Nature}\ }\textbf {\bibinfo {volume} {464}},\ \bibinfo {pages}
  {199} (\bibinfo {year} {2010})}\BibitemShut {NoStop}%
\bibitem [{\citenamefont {Chaloupka}\ \emph {et~al.}(2013)\citenamefont
  {Chaloupka}, \citenamefont {Jackeli},\ and\ \citenamefont
  {Khaliullin}}]{Khaliullin3}%
  \BibitemOpen
  \bibfield  {author} {\bibinfo {author} {\bibfnamefont
  {Ji\ifmmode\check{r}\else\v{r}\fi{}\'{i}}\ \bibnamefont {Chaloupka}},
  \bibinfo {author} {\bibfnamefont {George}\ \bibnamefont {Jackeli}}, \ and\
  \bibinfo {author} {\bibfnamefont {Giniyat}\ \bibnamefont {Khaliullin}},\
  }\bibfield  {title} {\enquote {\bibinfo {title} {\textit{Zigzag Magnetic
  Order in the Iridium Oxide} {N}a$_2${I}r{O}$_3$},}\ }\href {\doibase
  10.1103/PhysRevLett.110.097204} {\bibfield  {journal} {\bibinfo  {journal}
  {Phys. Rev. Lett.}\ }\textbf {\bibinfo {volume} {110}},\ \bibinfo {pages}
  {097204} (\bibinfo {year} {2013})}\BibitemShut {NoStop}%
\bibitem [{\citenamefont {Rau}\ \emph {et~al.}(2014)\citenamefont {Rau},
  \citenamefont {Lee},\ and\ \citenamefont {Kee}}]{KeePRL}%
  \BibitemOpen
  \bibfield  {author} {\bibinfo {author} {\bibfnamefont {Jeffrey~G.}\
  \bibnamefont {Rau}}, \bibinfo {author} {\bibfnamefont {Eric Kin-Ho}\
  \bibnamefont {Lee}}, \ and\ \bibinfo {author} {\bibfnamefont {Hae-Young}\
  \bibnamefont {Kee}},\ }\bibfield  {title} {\enquote {\bibinfo {title}
  {\textit{Generic Spin Model for the Honeycomb Iridates beyond the Kitaev
  Limit}},}\ }\href {\doibase 10.1103/PhysRevLett.112.077204} {\bibfield
  {journal} {\bibinfo  {journal} {Phys. Rev. Lett.}\ }\textbf {\bibinfo
  {volume} {112}},\ \bibinfo {pages} {077204} (\bibinfo {year}
  {2014})}\BibitemShut {NoStop}%
\bibitem [{\citenamefont {{Lou}}\ \emph {et~al.}(2015)\citenamefont {{Lou}},
  \citenamefont {{Liang}}, \citenamefont {{Yu}},\ and\ \citenamefont
  {{Chen}}}]{Chen1501}%
  \BibitemOpen
  \bibfield  {author} {\bibinfo {author} {\bibfnamefont {J.}~\bibnamefont
  {{Lou}}}, \bibinfo {author} {\bibfnamefont {L.}~\bibnamefont {{Liang}}},
  \bibinfo {author} {\bibfnamefont {Y.}~\bibnamefont {{Yu}}}, \ and\ \bibinfo
  {author} {\bibfnamefont {Y.}~\bibnamefont {{Chen}}},\ }\bibfield  {title}
  {\enquote {\bibinfo {title} {\textit{Global Phase Diagram of the Extended
  Kitaev-Heisenberg Model on Honeycomb Lattice}},}\ }\href@noop {} {\bibfield
  {journal} {\bibinfo  {journal} {arXiv:1501.06990}\ } (\bibinfo {year}
  {2015})}\BibitemShut {NoStop}%
\bibitem [{\citenamefont {Chaloupka}\ and\ \citenamefont
  {Khaliullin}(2015)}]{Khaliullin4}%
  \BibitemOpen
  \bibfield  {author} {\bibinfo {author} {\bibfnamefont
  {Ji\ifmmode\check{r}\else\v{r}\fi{}\'{\i}}\ \bibnamefont {Chaloupka}}\ and\
  \bibinfo {author} {\bibfnamefont {Giniyat}\ \bibnamefont {Khaliullin}},\
  }\bibfield  {title} {\enquote {\bibinfo {title} {\textit{Hidden Symmetries of
  the Extended Kitaev-Heisenberg Model: Implications for the Honeycomb-Lattice
  Iridates} {A}$_2${I}r{O}$_3$},}\ }\href {\doibase 10.1103/PhysRevB.92.024413}
  {\bibfield  {journal} {\bibinfo  {journal} {Phys. Rev. B}\ }\textbf {\bibinfo
  {volume} {92}},\ \bibinfo {pages} {024413} (\bibinfo {year}
  {2015})}\BibitemShut {NoStop}%
\bibitem [{\citenamefont {Tokura}\ and\ \citenamefont
  {Nagaosa}(2000)}]{Tokura21042000}%
  \BibitemOpen
  \bibfield  {author} {\bibinfo {author} {\bibfnamefont {Y.}~\bibnamefont
  {Tokura}}\ and\ \bibinfo {author} {\bibfnamefont {N.}~\bibnamefont
  {Nagaosa}},\ }\bibfield  {title} {\enquote {\bibinfo {title} {\textit{Orbital
  Physics in Transition-Metal Oxides}},}\ }\href {\doibase
  10.1126/science.288.5465.462} {\bibfield  {journal} {\bibinfo  {journal}
  {Science}\ }\textbf {\bibinfo {volume} {288}},\ \bibinfo {pages} {462}
  (\bibinfo {year} {2000})}\BibitemShut {NoStop}%
\bibitem [{\citenamefont {Wu}()}]{test1}%
  \BibitemOpen
  \bibfield  {author} {\bibinfo {author} {\bibfnamefont {Congjun}\ \bibnamefont
  {Wu}},\ }\href@noop {} {}\bibinfo {howpublished} {personal
  communication}\BibitemShut {NoStop}%
\bibitem [{\citenamefont {Wang}\ \emph {et~al.}(2011)\citenamefont {Wang},
  \citenamefont {Gu}, \citenamefont {Verstraete},\ and\ \citenamefont
  {Wen}}]{wang2011spin}%
  \BibitemOpen
  \bibfield  {author} {\bibinfo {author} {\bibfnamefont {Ling}\ \bibnamefont
  {Wang}}, \bibinfo {author} {\bibfnamefont {Zheng-Cheng}\ \bibnamefont {Gu}},
  \bibinfo {author} {\bibfnamefont {Frank}\ \bibnamefont {Verstraete}}, \ and\
  \bibinfo {author} {\bibfnamefont {Xiao-Gang}\ \bibnamefont {Wen}},\
  }\bibfield  {title} {\enquote {\bibinfo {title} {\textit{Spin-Liquid Phase in
  Spin-1/2 Square ${J}_1$-${J}_2$ Heisenberg Model: A Tensor Product State
  Approach}},}\ }\href@noop {} {\bibfield  {journal} {\bibinfo  {journal}
  {arXiv:1112.3331}\ } (\bibinfo {year} {2011})}\BibitemShut {NoStop}%
\bibitem [{\citenamefont {Harada}(2012)}]{HaradaMERA}%
  \BibitemOpen
  \bibfield  {author} {\bibinfo {author} {\bibfnamefont {Kenji}\ \bibnamefont
  {Harada}},\ }\bibfield  {title} {\enquote {\bibinfo {title}
  {\textit{Numerical Study of Incommensurability of the Spiral State on
  Spin-$\frac{1}{2}$ Spatially Anisotropic Triangular Antiferromagnets Using
  Entanglement Renormalization}},}\ }\href {\doibase
  10.1103/PhysRevB.86.184421} {\bibfield  {journal} {\bibinfo  {journal} {Phys.
  Rev. B}\ }\textbf {\bibinfo {volume} {86}},\ \bibinfo {pages} {184421}
  (\bibinfo {year} {2012})}\BibitemShut {NoStop}%
\bibitem [{\citenamefont {Corboz}\ and\ \citenamefont
  {Mila}(2013)}]{SSmodelTN}%
  \BibitemOpen
  \bibfield  {author} {\bibinfo {author} {\bibfnamefont {Philippe}\
  \bibnamefont {Corboz}}\ and\ \bibinfo {author} {\bibfnamefont {Fr\'ed\'eric}\
  \bibnamefont {Mila}},\ }\bibfield  {title} {\enquote {\bibinfo {title}
  {\textit{Tensor Network Study of the Shastry-Sutherland Model in Zero
  Magnetic Field}},}\ }\href {\doibase 10.1103/PhysRevB.87.115144} {\bibfield
  {journal} {\bibinfo  {journal} {Phys. Rev. B}\ }\textbf {\bibinfo {volume}
  {87}},\ \bibinfo {pages} {115144} (\bibinfo {year} {2013})}\BibitemShut
  {NoStop}%
\bibitem [{\citenamefont {Xie}\ \emph {et~al.}(2014)\citenamefont {Xie},
  \citenamefont {Chen}, \citenamefont {Yu}, \citenamefont {Kong}, \citenamefont
  {Normand},\ and\ \citenamefont {Xiang}}]{XiePRX14}%
  \BibitemOpen
  \bibfield  {author} {\bibinfo {author} {\bibfnamefont {Z.~Y.}\ \bibnamefont
  {Xie}}, \bibinfo {author} {\bibfnamefont {J.}~\bibnamefont {Chen}}, \bibinfo
  {author} {\bibfnamefont {J.~F.}\ \bibnamefont {Yu}}, \bibinfo {author}
  {\bibfnamefont {X.}~\bibnamefont {Kong}}, \bibinfo {author} {\bibfnamefont
  {B.}~\bibnamefont {Normand}}, \ and\ \bibinfo {author} {\bibfnamefont
  {T.}~\bibnamefont {Xiang}},\ }\bibfield  {title} {\enquote {\bibinfo {title}
  {\textit{Tensor Renormalization of Quantum Many-Body Systems Using Projected
  Entangled Simplex States}},}\ }\href {\doibase 10.1103/PhysRevX.4.011025}
  {\bibfield  {journal} {\bibinfo  {journal} {Phys. Rev. X}\ }\textbf {\bibinfo
  {volume} {4}},\ \bibinfo {pages} {011025} (\bibinfo {year}
  {2014})}\BibitemShut {NoStop}%
\bibitem [{\citenamefont {Levin}\ and\ \citenamefont {Nave}(2007)}]{LevinTRG}%
  \BibitemOpen
  \bibfield  {author} {\bibinfo {author} {\bibfnamefont {Michael}\ \bibnamefont
  {Levin}}\ and\ \bibinfo {author} {\bibfnamefont {Cody~P.}\ \bibnamefont
  {Nave}},\ }\bibfield  {title} {\enquote {\bibinfo {title} {\textit{Tensor
  Renormalization Group Approach to Two-Dimensional Classical Lattice
  Models}},}\ }\href {\doibase 10.1103/PhysRevLett.99.120601} {\bibfield
  {journal} {\bibinfo  {journal} {Phys. Rev. Lett.}\ }\textbf {\bibinfo
  {volume} {99}},\ \bibinfo {pages} {120601} (\bibinfo {year}
  {2007})}\BibitemShut {NoStop}%
\bibitem [{\citenamefont {Xie}\ \emph {et~al.}(2012)\citenamefont {Xie},
  \citenamefont {Chen}, \citenamefont {Qin}, \citenamefont {Zhu}, \citenamefont
  {Yang},\ and\ \citenamefont {Xiang}}]{HOTRG1}%
  \BibitemOpen
  \bibfield  {author} {\bibinfo {author} {\bibfnamefont {Z.~Y.}\ \bibnamefont
  {Xie}}, \bibinfo {author} {\bibfnamefont {J.}~\bibnamefont {Chen}}, \bibinfo
  {author} {\bibfnamefont {M.~P.}\ \bibnamefont {Qin}}, \bibinfo {author}
  {\bibfnamefont {J.~W.}\ \bibnamefont {Zhu}}, \bibinfo {author} {\bibfnamefont
  {L.~P.}\ \bibnamefont {Yang}}, \ and\ \bibinfo {author} {\bibfnamefont
  {T.}~\bibnamefont {Xiang}},\ }\bibfield  {title} {\enquote {\bibinfo {title}
  {\textit{Coarse-Graining Renormalization by Higher-Order Singular Value
  Decomposition}},}\ }\href {\doibase 10.1103/PhysRevB.86.045139} {\bibfield
  {journal} {\bibinfo  {journal} {Phys. Rev. B}\ }\textbf {\bibinfo {volume}
  {86}},\ \bibinfo {pages} {045139} (\bibinfo {year} {2012})}\BibitemShut
  {NoStop}%
\bibitem [{not()}]{noteSL}%
  \BibitemOpen
  \href@noop {} {}\bibinfo {note} {In this paper, we use the term ``spin
  liquid" to denote a phase that shows no long-range spin order according to
  our tensor network algorithms. At the special point, $\theta'=90^\circ$ the
  tripod model reduces to the Kitaev model and its ground state is well
  established to be a gapless spin liquid. Our numerical results offer evidence
  that the same spin-liquid state will survive away from the Kitaev point. The
  nature of the excitations (e.g. their fractional statistics) remains to be
  checked in order to firmly establish that it is a quantum spin
  liquid.}\BibitemShut {Stop}%
\bibitem [{\citenamefont {{Verstraete}}\ and\ \citenamefont
  {{Cirac}}(2004)}]{PEPS1}%
  \BibitemOpen
  \bibfield  {author} {\bibinfo {author} {\bibfnamefont {F.}~\bibnamefont
  {{Verstraete}}}\ and\ \bibinfo {author} {\bibfnamefont {J.~I.}\ \bibnamefont
  {{Cirac}}},\ }\bibfield  {title} {\enquote {\bibinfo {title}
  {\textit{Renormalization algorithms for Quantum-Many Body Systems in two and
  higher dimensions}},}\ }\href@noop {} {\bibfield  {journal} {\bibinfo
  {journal} {arXiv:cond-mat/0407066}\ } (\bibinfo {year} {2004})}\BibitemShut
  {NoStop}%
\bibitem [{\citenamefont {Verstraete}\ \emph {et~al.}(2006)\citenamefont
  {Verstraete}, \citenamefont {Wolf}, \citenamefont {Perez-Garcia},\ and\
  \citenamefont {Cirac}}]{PEPS2}%
  \BibitemOpen
  \bibfield  {author} {\bibinfo {author} {\bibfnamefont {F.}~\bibnamefont
  {Verstraete}}, \bibinfo {author} {\bibfnamefont {M.~M.}\ \bibnamefont
  {Wolf}}, \bibinfo {author} {\bibfnamefont {D.}~\bibnamefont {Perez-Garcia}},
  \ and\ \bibinfo {author} {\bibfnamefont {J.~I.}\ \bibnamefont {Cirac}},\
  }\bibfield  {title} {\enquote {\bibinfo {title} {\textit{Criticality, the
  Area Law, and the Computational Power of Projected Entangled Pair States}},}\
  }\href {\doibase 10.1103/PhysRevLett.96.220601} {\bibfield  {journal}
  {\bibinfo  {journal} {Phys. Rev. Lett.}\ }\textbf {\bibinfo {volume} {96}},\
  \bibinfo {pages} {220601} (\bibinfo {year} {2006})}\BibitemShut {NoStop}%
\bibitem [{\citenamefont {Or\'us}(2014)}]{Orus2014}%
  \BibitemOpen
  \bibfield  {author} {\bibinfo {author} {\bibfnamefont {Román}\ \bibnamefont
  {Or\'us}},\ }\bibfield  {title} {\enquote {\bibinfo {title} {\textit{A
  Practical Introduction to Tensor Networks: Matrix Product States and
  Projected Entangled Pair States}},}\ }\href@noop {} {\bibfield  {journal}
  {\bibinfo  {journal} {Annals of Physics}\ }\textbf {\bibinfo {volume}
  {349}},\ \bibinfo {pages} {117} (\bibinfo {year} {2014})}\BibitemShut
  {NoStop}%
\bibitem [{\citenamefont {Jordan}\ \emph {et~al.}(2008)\citenamefont {Jordan},
  \citenamefont {Or\'us}, \citenamefont {Vidal}, \citenamefont {Verstraete},\
  and\ \citenamefont {Cirac}}]{iPEPS}%
  \BibitemOpen
  \bibfield  {author} {\bibinfo {author} {\bibfnamefont {J.}~\bibnamefont
  {Jordan}}, \bibinfo {author} {\bibfnamefont {R.}~\bibnamefont {Or\'us}},
  \bibinfo {author} {\bibfnamefont {G.}~\bibnamefont {Vidal}}, \bibinfo
  {author} {\bibfnamefont {F.}~\bibnamefont {Verstraete}}, \ and\ \bibinfo
  {author} {\bibfnamefont {J.~I.}\ \bibnamefont {Cirac}},\ }\bibfield  {title}
  {\enquote {\bibinfo {title} {\textit{Classical Simulation of Infinite-Size
  Quantum Lattice Systems in Two Spatial Dimensions}},}\ }\href {\doibase
  10.1103/PhysRevLett.101.250602} {\bibfield  {journal} {\bibinfo  {journal}
  {Phys. Rev. Lett.}\ }\textbf {\bibinfo {volume} {101}},\ \bibinfo {pages}
  {250602} (\bibinfo {year} {2008})}\BibitemShut {NoStop}%
\bibitem [{\citenamefont {Corboz}\ \emph {et~al.}(2010)\citenamefont {Corboz},
  \citenamefont {Or\'us}, \citenamefont {Bauer},\ and\ \citenamefont
  {Vidal}}]{iPEPS2}%
  \BibitemOpen
  \bibfield  {author} {\bibinfo {author} {\bibfnamefont {Philippe}\
  \bibnamefont {Corboz}}, \bibinfo {author} {\bibfnamefont {Rom\'an}\
  \bibnamefont {Or\'us}}, \bibinfo {author} {\bibfnamefont {Bela}\ \bibnamefont
  {Bauer}}, \ and\ \bibinfo {author} {\bibfnamefont {Guifr\'e}\ \bibnamefont
  {Vidal}},\ }\bibfield  {title} {\enquote {\bibinfo {title}
  {\textit{Simulation of Strongly Correlated Fermions in Two Spatial Dimensions
  with Fermionic Projected Entangled-Pair States}},}\ }\href {\doibase
  10.1103/PhysRevB.81.165104} {\bibfield  {journal} {\bibinfo  {journal} {Phys.
  Rev. B}\ }\textbf {\bibinfo {volume} {81}},\ \bibinfo {pages} {165104}
  (\bibinfo {year} {2010})}\BibitemShut {NoStop}%
\bibitem [{\citenamefont {Corboz}\ \emph {et~al.}(2011)\citenamefont {Corboz},
  \citenamefont {White}, \citenamefont {Vidal},\ and\ \citenamefont
  {Troyer}}]{tJ_Troyer11}%
  \BibitemOpen
  \bibfield  {author} {\bibinfo {author} {\bibfnamefont {Philippe}\
  \bibnamefont {Corboz}}, \bibinfo {author} {\bibfnamefont {Steven~R.}\
  \bibnamefont {White}}, \bibinfo {author} {\bibfnamefont {Guifr\'e}\
  \bibnamefont {Vidal}}, \ and\ \bibinfo {author} {\bibfnamefont {Matthias}\
  \bibnamefont {Troyer}},\ }\bibfield  {title} {\enquote {\bibinfo {title}
  {\textit{Stripes in the Two-Dimensional $t$-${J}$ Model with Infinite
  Projected Entangled-Pair States}},}\ }\href {\doibase
  10.1103/PhysRevB.84.041108} {\bibfield  {journal} {\bibinfo  {journal} {Phys.
  Rev. B}\ }\textbf {\bibinfo {volume} {84}},\ \bibinfo {pages} {041108}
  (\bibinfo {year} {2011})}\BibitemShut {NoStop}%
\bibitem [{\citenamefont {Corboz}\ \emph {et~al.}(2014)\citenamefont {Corboz},
  \citenamefont {Rice},\ and\ \citenamefont {Troyer}}]{tJ_Troyer14}%
  \BibitemOpen
  \bibfield  {author} {\bibinfo {author} {\bibfnamefont {Philippe}\
  \bibnamefont {Corboz}}, \bibinfo {author} {\bibfnamefont {T.~M.}\
  \bibnamefont {Rice}}, \ and\ \bibinfo {author} {\bibfnamefont {Matthias}\
  \bibnamefont {Troyer}},\ }\bibfield  {title} {\enquote {\bibinfo {title}
  {\textit{Competing States in the $t$-{$J$} Model: Uniform $d$-Wave State
  versus Stripe State}},}\ }\href {\doibase 10.1103/PhysRevLett.113.046402}
  {\bibfield  {journal} {\bibinfo  {journal} {Phys. Rev. Lett.}\ }\textbf
  {\bibinfo {volume} {113}},\ \bibinfo {pages} {046402} (\bibinfo {year}
  {2014})}\BibitemShut {NoStop}%
\bibitem [{\citenamefont {Jiang}\ \emph {et~al.}(2008)\citenamefont {Jiang},
  \citenamefont {Weng},\ and\ \citenamefont {Xiang}}]{simpleup}%
  \BibitemOpen
  \bibfield  {author} {\bibinfo {author} {\bibfnamefont {H.~C.}\ \bibnamefont
  {Jiang}}, \bibinfo {author} {\bibfnamefont {Z.~Y.}\ \bibnamefont {Weng}}, \
  and\ \bibinfo {author} {\bibfnamefont {T.}~\bibnamefont {Xiang}},\ }\bibfield
   {title} {\enquote {\bibinfo {title} {\textit{Accurate Determination of
  Tensor Network State of Quantum Lattice Models in Two Dimensions}},}\ }\href
  {\doibase 10.1103/PhysRevLett.101.090603} {\bibfield  {journal} {\bibinfo
  {journal} {Phys. Rev. Lett.}\ }\textbf {\bibinfo {volume} {101}},\ \bibinfo
  {pages} {090603} (\bibinfo {year} {2008})}\BibitemShut {NoStop}%
\bibitem [{\citenamefont {Vidal}(2003)}]{TEBD}%
  \BibitemOpen
  \bibfield  {author} {\bibinfo {author} {\bibfnamefont {Guifr\'e}\
  \bibnamefont {Vidal}},\ }\bibfield  {title} {\enquote {\bibinfo {title}
  {\textit{Efficient Classical Simulation of Slightly Entangled Quantum
  Computations}},}\ }\href {\doibase 10.1103/PhysRevLett.91.147902} {\bibfield
  {journal} {\bibinfo  {journal} {Phys. Rev. Lett.}\ }\textbf {\bibinfo
  {volume} {91}},\ \bibinfo {pages} {147902} (\bibinfo {year}
  {2003})}\BibitemShut {NoStop}%
\bibitem [{\citenamefont {Nasu}\ \emph {et~al.}(2008)\citenamefont {Nasu},
  \citenamefont {Nagano}, \citenamefont {Naka},\ and\ \citenamefont
  {Ishihara}}]{nasu}%
  \BibitemOpen
  \bibfield  {author} {\bibinfo {author} {\bibfnamefont {J.}~\bibnamefont
  {Nasu}}, \bibinfo {author} {\bibfnamefont {A.}~\bibnamefont {Nagano}},
  \bibinfo {author} {\bibfnamefont {M.}~\bibnamefont {Naka}}, \ and\ \bibinfo
  {author} {\bibfnamefont {S.}~\bibnamefont {Ishihara}},\ }\bibfield  {title}
  {\enquote {\bibinfo {title} {\textit{Doubly Degenerate Orbital System in
  Honeycomb Lattice: Implication of Orbital State in Layered Iron Oxide}},}\
  }\href {\doibase 10.1103/PhysRevB.78.024416} {\bibfield  {journal} {\bibinfo
  {journal} {Phys. Rev. B}\ }\textbf {\bibinfo {volume} {78}},\ \bibinfo
  {pages} {024416} (\bibinfo {year} {2008})}\BibitemShut {NoStop}%
\bibitem [{\citenamefont {Murthy}\ \emph {et~al.}(1997)\citenamefont {Murthy},
  \citenamefont {Arovas},\ and\ \citenamefont {Auerbach}}]{HP1}%
  \BibitemOpen
  \bibfield  {author} {\bibinfo {author} {\bibfnamefont {Ganpathy}\
  \bibnamefont {Murthy}}, \bibinfo {author} {\bibfnamefont {Daniel}\
  \bibnamefont {Arovas}}, \ and\ \bibinfo {author} {\bibfnamefont {Assa}\
  \bibnamefont {Auerbach}},\ }\bibfield  {title} {\enquote {\bibinfo {title}
  {\textit{Superfluids and Supersolids on Frustrated Two-Dimensional
  Lattices}},}\ }\href {\doibase 10.1103/PhysRevB.55.3104} {\bibfield
  {journal} {\bibinfo  {journal} {Phys. Rev. B}\ }\textbf {\bibinfo {volume}
  {55}},\ \bibinfo {pages} {3104} (\bibinfo {year} {1997})}\BibitemShut
  {NoStop}%
\bibitem [{\citenamefont {Duan}\ \emph {et~al.}(2003)\citenamefont {Duan},
  \citenamefont {Demler},\ and\ \citenamefont {Lukin}}]{Duan}%
  \BibitemOpen
  \bibfield  {author} {\bibinfo {author} {\bibfnamefont {L.-M.}\ \bibnamefont
  {Duan}}, \bibinfo {author} {\bibfnamefont {E.}~\bibnamefont {Demler}}, \ and\
  \bibinfo {author} {\bibfnamefont {M.~D.}\ \bibnamefont {Lukin}},\ }\bibfield
  {title} {\enquote {\bibinfo {title} {\textit{Controlling Spin Exchange
  Interactions of Ultracold Atoms in Optical Lattices}},}\ }\href {\doibase
  10.1103/PhysRevLett.91.090402} {\bibfield  {journal} {\bibinfo  {journal}
  {Phys. Rev. Lett.}\ }\textbf {\bibinfo {volume} {91}},\ \bibinfo {pages}
  {090402} (\bibinfo {year} {2003})}\BibitemShut {NoStop}%
\bibitem [{\citenamefont {Liu}\ \emph {et~al.}(2004)\citenamefont {Liu},
  \citenamefont {Wilczek},\ and\ \citenamefont {Zoller}}]{WVLiu04}%
  \BibitemOpen
  \bibfield  {author} {\bibinfo {author} {\bibfnamefont {W.~Vincent}\
  \bibnamefont {Liu}}, \bibinfo {author} {\bibfnamefont {Frank}\ \bibnamefont
  {Wilczek}}, \ and\ \bibinfo {author} {\bibfnamefont {Peter}\ \bibnamefont
  {Zoller}},\ }\bibfield  {title} {\enquote {\bibinfo {title}
  {\textit{Spin-Dependent Hubbard Model and a Quantum Phase Transition in Cold
  Atoms}},}\ }\href {\doibase 10.1103/PhysRevA.70.033603} {\bibfield  {journal}
  {\bibinfo  {journal} {Phys. Rev. A}\ }\textbf {\bibinfo {volume} {70}},\
  \bibinfo {pages} {033603} (\bibinfo {year} {2004})}\BibitemShut {NoStop}%
\bibitem [{\citenamefont {Yi}\ \emph {et~al.}(2008)\citenamefont {Yi},
  \citenamefont {Daley}, \citenamefont {Pupillo},\ and\ \citenamefont
  {Zoller}}]{PZoller08}%
  \BibitemOpen
  \bibfield  {author} {\bibinfo {author} {\bibfnamefont {W}~\bibnamefont {Yi}},
  \bibinfo {author} {\bibfnamefont {A~J}\ \bibnamefont {Daley}}, \bibinfo
  {author} {\bibfnamefont {G}~\bibnamefont {Pupillo}}, \ and\ \bibinfo {author}
  {\bibfnamefont {P}~\bibnamefont {Zoller}},\ }\bibfield  {title} {\enquote
  {\bibinfo {title} {\textit{State-Dependent, Addressable Subwavelength
  Lattices with Cold Atoms}},}\ }\href
  {http://stacks.iop.org/1367-2630/10/i=7/a=073015} {\bibfield  {journal}
  {\bibinfo  {journal} {New Journal of Physics}\ }\textbf {\bibinfo {volume}
  {10}},\ \bibinfo {pages} {073015} (\bibinfo {year} {2008})}\BibitemShut
  {NoStop}%
\bibitem [{\citenamefont {McKay}\ and\ \citenamefont
  {DeMarco}(2010)}]{DeMarco10}%
  \BibitemOpen
  \bibfield  {author} {\bibinfo {author} {\bibfnamefont {D}~\bibnamefont
  {McKay}}\ and\ \bibinfo {author} {\bibfnamefont {B}~\bibnamefont {DeMarco}},\
  }\bibfield  {title} {\enquote {\bibinfo {title} {\textit{Thermometry with
  Spin-Dependent Lattices}},}\ }\href
  {http://stacks.iop.org/1367-2630/12/i=5/a=055013} {\bibfield  {journal}
  {\bibinfo  {journal} {New Journal of Physics}\ }\textbf {\bibinfo {volume}
  {12}},\ \bibinfo {pages} {055013} (\bibinfo {year} {2010})}\BibitemShut
  {NoStop}%
\bibitem [{\citenamefont {Daley}\ \emph {et~al.}(2011)\citenamefont {Daley},
  \citenamefont {Ye},\ and\ \citenamefont {Zoller}}]{PZoller11}%
  \BibitemOpen
  \bibfield  {author} {\bibinfo {author} {\bibfnamefont {A.J.}\ \bibnamefont
  {Daley}}, \bibinfo {author} {\bibfnamefont {J.}~\bibnamefont {Ye}}, \ and\
  \bibinfo {author} {\bibfnamefont {P.}~\bibnamefont {Zoller}},\ }\bibfield
  {title} {\enquote {\bibinfo {title} {\textit{State-Dependent Lattices for
  Quantum Computing with Alkaline-Earth-Metal Atoms}},}\ }\href {\doibase
  10.1140/epjd/e2011-20095-2} {\bibfield  {journal} {\bibinfo  {journal} {The
  European Physical Journal D}\ }\textbf {\bibinfo {volume} {65}},\ \bibinfo
  {pages} {207} (\bibinfo {year} {2011})}\BibitemShut {NoStop}%
\bibitem [{\citenamefont {Lee}\ \emph {et~al.}(2007)\citenamefont {Lee},
  \citenamefont {Anderlini}, \citenamefont {Brown}, \citenamefont
  {Sebby-Strabley}, \citenamefont {Phillips},\ and\ \citenamefont
  {Porto}}]{Porto07}%
  \BibitemOpen
  \bibfield  {author} {\bibinfo {author} {\bibfnamefont {P.~J.}\ \bibnamefont
  {Lee}}, \bibinfo {author} {\bibfnamefont {M.}~\bibnamefont {Anderlini}},
  \bibinfo {author} {\bibfnamefont {B.~L.}\ \bibnamefont {Brown}}, \bibinfo
  {author} {\bibfnamefont {J.}~\bibnamefont {Sebby-Strabley}}, \bibinfo
  {author} {\bibfnamefont {W.~D.}\ \bibnamefont {Phillips}}, \ and\ \bibinfo
  {author} {\bibfnamefont {J.~V.}\ \bibnamefont {Porto}},\ }\bibfield  {title}
  {\enquote {\bibinfo {title} {\textit{Sublattice Addressing and Spin-Dependent
  Motion of Atoms in a Double-Well Lattice}},}\ }\href {\doibase
  10.1103/PhysRevLett.99.020402} {\bibfield  {journal} {\bibinfo  {journal}
  {Phys. Rev. Lett.}\ }\textbf {\bibinfo {volume} {99}},\ \bibinfo {pages}
  {020402} (\bibinfo {year} {2007})}\BibitemShut {NoStop}%
\bibitem [{\citenamefont {Soltan-Panahi}\ \emph {et~al.}(2011)\citenamefont
  {Soltan-Panahi}, \citenamefont {Struck}, \citenamefont {Hauke}, \citenamefont
  {Bick}, \citenamefont {Plenkers}, \citenamefont {Meineke}, \citenamefont
  {Becker}, \citenamefont {Windpassinger}, \citenamefont {Lewenstein},\ and\
  \citenamefont {Sengstock}}]{Lewenstein11}%
  \BibitemOpen
  \bibfield  {author} {\bibinfo {author} {\bibfnamefont {P.}~\bibnamefont
  {Soltan-Panahi}}, \bibinfo {author} {\bibfnamefont {J.}~\bibnamefont
  {Struck}}, \bibinfo {author} {\bibfnamefont {P.}~\bibnamefont {Hauke}},
  \bibinfo {author} {\bibfnamefont {A.}~\bibnamefont {Bick}}, \bibinfo {author}
  {\bibfnamefont {W.}~\bibnamefont {Plenkers}}, \bibinfo {author}
  {\bibfnamefont {G.}~\bibnamefont {Meineke}}, \bibinfo {author} {\bibfnamefont
  {C.}~\bibnamefont {Becker}}, \bibinfo {author} {\bibfnamefont
  {P.}~\bibnamefont {Windpassinger}}, \bibinfo {author} {\bibfnamefont
  {M.}~\bibnamefont {Lewenstein}}, \ and\ \bibinfo {author} {\bibfnamefont
  {K.}~\bibnamefont {Sengstock}},\ }\bibfield  {title} {\enquote {\bibinfo
  {title} {\textit{Multi-Component Quantum Gases in Spin-Dependent Hexagonal
  Lattices}},}\ }\href {http://dx.doi.org/10.1038/nphys1916} {\bibfield
  {journal} {\bibinfo  {journal} {Nat Phys}\ }\textbf {\bibinfo {volume} {7}},\
  \bibinfo {pages} {434} (\bibinfo {year} {2011})}\BibitemShut {NoStop}%
\bibitem [{\citenamefont {Jotzu}\ \emph {et~al.}(2014)\citenamefont {Jotzu},
  \citenamefont {Messer}, \citenamefont {Desbuquois}, \citenamefont {Lebrat},
  \citenamefont {Uehlinger}, \citenamefont {Greif},\ and\ \citenamefont
  {Esslinger}}]{EsslingerNat14}%
  \BibitemOpen
  \bibfield  {author} {\bibinfo {author} {\bibfnamefont {Gregor}\ \bibnamefont
  {Jotzu}}, \bibinfo {author} {\bibfnamefont {Michael}\ \bibnamefont {Messer}},
  \bibinfo {author} {\bibfnamefont {Remi}\ \bibnamefont {Desbuquois}}, \bibinfo
  {author} {\bibfnamefont {Martin}\ \bibnamefont {Lebrat}}, \bibinfo {author}
  {\bibfnamefont {Thomas}\ \bibnamefont {Uehlinger}}, \bibinfo {author}
  {\bibfnamefont {Daniel}\ \bibnamefont {Greif}}, \ and\ \bibinfo {author}
  {\bibfnamefont {Tilman}\ \bibnamefont {Esslinger}},\ }\bibfield  {title}
  {\enquote {\bibinfo {title} {\textit{Experimental Realization of the
  Topological Haldane Model with Ultracold Fermions}},}\ }\href
  {http://dx.doi.org/10.1038/nature13915} {\bibfield  {journal} {\bibinfo
  {journal} {Nature}\ }\textbf {\bibinfo {volume} {515}},\ \bibinfo {pages}
  {237} (\bibinfo {year} {2014})}\BibitemShut {NoStop}%
\bibitem [{\citenamefont {Jotzu}\ \emph {et~al.}(2015)\citenamefont {Jotzu},
  \citenamefont {Messer}, \citenamefont {G\"org}, \citenamefont {Greif},
  \citenamefont {Desbuquois},\ and\ \citenamefont {Esslinger}}]{Esslinger}%
  \BibitemOpen
  \bibfield  {author} {\bibinfo {author} {\bibfnamefont {Gregor}\ \bibnamefont
  {Jotzu}}, \bibinfo {author} {\bibfnamefont {Michael}\ \bibnamefont {Messer}},
  \bibinfo {author} {\bibfnamefont {Frederik}\ \bibnamefont {G\"org}}, \bibinfo
  {author} {\bibfnamefont {Daniel}\ \bibnamefont {Greif}}, \bibinfo {author}
  {\bibfnamefont {R\'emi}\ \bibnamefont {Desbuquois}}, \ and\ \bibinfo {author}
  {\bibfnamefont {Tilman}\ \bibnamefont {Esslinger}},\ }\bibfield  {title}
  {\enquote {\bibinfo {title} {\textit{Creating State-Dependent Lattices for
  Ultracold Fermions by Magnetic Gradient Modulation}},}\ }\href {\doibase
  10.1103/PhysRevLett.115.073002} {\bibfield  {journal} {\bibinfo  {journal}
  {Phys. Rev. Lett.}\ }\textbf {\bibinfo {volume} {115}},\ \bibinfo {pages}
  {073002} (\bibinfo {year} {2015})}\BibitemShut {NoStop}%
\bibitem [{\citenamefont {Xu}\ \emph {et~al.}(2013)\citenamefont {Xu},
  \citenamefont {You},\ and\ \citenamefont {Ueda}}]{Xupra}%
  \BibitemOpen
  \bibfield  {author} {\bibinfo {author} {\bibfnamefont {Zhi-Fang}\
  \bibnamefont {Xu}}, \bibinfo {author} {\bibfnamefont {Li}~\bibnamefont
  {You}}, \ and\ \bibinfo {author} {\bibfnamefont {Masahito}\ \bibnamefont
  {Ueda}},\ }\bibfield  {title} {\enquote {\bibinfo {title} {\textit{Atomic
  Spin-Orbit Coupling Synthesized with Magnetic-Field-Gradient Pulses}},}\
  }\href {\doibase 10.1103/PhysRevA.87.063634} {\bibfield  {journal} {\bibinfo
  {journal} {Phys. Rev. A}\ }\textbf {\bibinfo {volume} {87}},\ \bibinfo
  {pages} {063634} (\bibinfo {year} {2013})}\BibitemShut {NoStop}%
\bibitem [{\citenamefont {Anderson}\ \emph {et~al.}(2013)\citenamefont
  {Anderson}, \citenamefont {Spielman},\ and\ \citenamefont
  {Juzeli\ifmmode~\bar{u}\else \={u}\fi{}nas}}]{MGM2}%
  \BibitemOpen
  \bibfield  {author} {\bibinfo {author} {\bibfnamefont {Brandon~M.}\
  \bibnamefont {Anderson}}, \bibinfo {author} {\bibfnamefont {I.~B.}\
  \bibnamefont {Spielman}}, \ and\ \bibinfo {author} {\bibfnamefont
  {Gediminas}\ \bibnamefont {Juzeli\ifmmode~\bar{u}\else \={u}\fi{}nas}},\
  }\bibfield  {title} {\enquote {\bibinfo {title} {\textit{Magnetically
  Generated Spin-Orbit Coupling for Ultracold Atoms}},}\ }\href {\doibase
  10.1103/PhysRevLett.111.125301} {\bibfield  {journal} {\bibinfo  {journal}
  {Phys. Rev. Lett.}\ }\textbf {\bibinfo {volume} {111}},\ \bibinfo {pages}
  {125301} (\bibinfo {year} {2013})}\BibitemShut {NoStop}%
\bibitem [{\citenamefont {You}\ \emph {et~al.}(2010)\citenamefont {You},
  \citenamefont {Shi}, \citenamefont {Hu},\ and\ \citenamefont
  {Nori}}]{SCcircuit}%
  \BibitemOpen
  \bibfield  {author} {\bibinfo {author} {\bibfnamefont {J.~Q.}\ \bibnamefont
  {You}}, \bibinfo {author} {\bibfnamefont {Xiao-Feng}\ \bibnamefont {Shi}},
  \bibinfo {author} {\bibfnamefont {Xuedong}\ \bibnamefont {Hu}}, \ and\
  \bibinfo {author} {\bibfnamefont {Franco}\ \bibnamefont {Nori}},\ }\bibfield
  {title} {\enquote {\bibinfo {title} {\textit{Quantum Emulation of a Spin
  System with Topologically Protected Ground States Using Superconducting
  Quantum Circuits}},}\ }\href {\doibase 10.1103/PhysRevB.81.014505} {\bibfield
   {journal} {\bibinfo  {journal} {Phys. Rev. B}\ }\textbf {\bibinfo {volume}
  {81}},\ \bibinfo {pages} {014505} (\bibinfo {year} {2010})}\BibitemShut
  {NoStop}%
\bibitem [{\citenamefont {Nishino}\ and\ \citenamefont
  {Okunishi}(1996)}]{CTM1}%
  \BibitemOpen
  \bibfield  {author} {\bibinfo {author} {\bibfnamefont {Tomotoshi}\
  \bibnamefont {Nishino}}\ and\ \bibinfo {author} {\bibfnamefont {Kouichi}\
  \bibnamefont {Okunishi}},\ }\bibfield  {title} {\enquote {\bibinfo {title}
  {\textit{Corner Transfer Matrix Renormalization Group Method}},}\ }\href
  {\doibase 10.1143/JPSJ.65.891} {\bibfield  {journal} {\bibinfo  {journal}
  {Journal of the Physical Society of Japan}\ }\textbf {\bibinfo {volume}
  {65}},\ \bibinfo {pages} {891} (\bibinfo {year} {1996})}\BibitemShut
  {NoStop}%
\bibitem [{\citenamefont {Or\'us}\ and\ \citenamefont {Vidal}(2009)}]{CTM2}%
  \BibitemOpen
  \bibfield  {author} {\bibinfo {author} {\bibfnamefont {Rom\'an}\ \bibnamefont
  {Or\'us}}\ and\ \bibinfo {author} {\bibfnamefont {Guifr\'e}\ \bibnamefont
  {Vidal}},\ }\bibfield  {title} {\enquote {\bibinfo {title}
  {\textit{Simulation of Two-Dimensional Quantum Systems on an Infinite Lattice
  Revisited: Corner Transfer Matrix for Tensor Contraction}},}\ }\href
  {\doibase 10.1103/PhysRevB.80.094403} {\bibfield  {journal} {\bibinfo
  {journal} {Phys. Rev. B}\ }\textbf {\bibinfo {volume} {80}},\ \bibinfo
  {pages} {094403} (\bibinfo {year} {2009})}\BibitemShut {NoStop}%
\bibitem [{\citenamefont {Osorio~Iregui}\ \emph {et~al.}(2014)\citenamefont
  {Osorio~Iregui}, \citenamefont {Corboz},\ and\ \citenamefont
  {Troyer}}]{iPEPSKH}%
  \BibitemOpen
  \bibfield  {author} {\bibinfo {author} {\bibfnamefont {Juan}\ \bibnamefont
  {Osorio~Iregui}}, \bibinfo {author} {\bibfnamefont {Philippe}\ \bibnamefont
  {Corboz}}, \ and\ \bibinfo {author} {\bibfnamefont {Matthias}\ \bibnamefont
  {Troyer}},\ }\bibfield  {title} {\enquote {\bibinfo {title} {\textit{Probing
  the Stability of the Spin-Liquid Phases in the Kitaev-Heisenberg Model Using
  Tensor Network Algorithms}},}\ }\href {\doibase 10.1103/PhysRevB.90.195102}
  {\bibfield  {journal} {\bibinfo  {journal} {Phys. Rev. B}\ }\textbf {\bibinfo
  {volume} {90}},\ \bibinfo {pages} {195102} (\bibinfo {year}
  {2014})}\BibitemShut {NoStop}%
\bibitem [{\citenamefont {Corboz}\ \emph {et~al.}(2012)\citenamefont {Corboz},
  \citenamefont {Lajk\'o}, \citenamefont {L\"auchli}, \citenamefont {Penc},\
  and\ \citenamefont {Mila}}]{KKprx}%
  \BibitemOpen
  \bibfield  {author} {\bibinfo {author} {\bibfnamefont {Philippe}\
  \bibnamefont {Corboz}}, \bibinfo {author} {\bibfnamefont {Mikl\'os}\
  \bibnamefont {Lajk\'o}}, \bibinfo {author} {\bibfnamefont {Andreas~M.}\
  \bibnamefont {L\"auchli}}, \bibinfo {author} {\bibfnamefont {Karlo}\
  \bibnamefont {Penc}}, \ and\ \bibinfo {author} {\bibfnamefont {Fr\'ed\'eric}\
  \bibnamefont {Mila}},\ }\bibfield  {title} {\enquote {\bibinfo {title}
  {\textit{Spin-Orbital Quantum Liquid on the Honeycomb Lattice}},}\ }\href
  {\doibase 10.1103/PhysRevX.2.041013} {\bibfield  {journal} {\bibinfo
  {journal} {Phys. Rev. X}\ }\textbf {\bibinfo {volume} {2}},\ \bibinfo {pages}
  {041013} (\bibinfo {year} {2012})}\BibitemShut {NoStop}%
\bibitem [{\citenamefont {Xie}\ \emph {et~al.}(2009)\citenamefont {Xie},
  \citenamefont {Jiang}, \citenamefont {Chen}, \citenamefont {Weng},\ and\
  \citenamefont {Xiang}}]{SRG}%
  \BibitemOpen
  \bibfield  {author} {\bibinfo {author} {\bibfnamefont {Z.~Y.}\ \bibnamefont
  {Xie}}, \bibinfo {author} {\bibfnamefont {H.~C.}\ \bibnamefont {Jiang}},
  \bibinfo {author} {\bibfnamefont {Q.~N.}\ \bibnamefont {Chen}}, \bibinfo
  {author} {\bibfnamefont {Z.~Y.}\ \bibnamefont {Weng}}, \ and\ \bibinfo
  {author} {\bibfnamefont {T.}~\bibnamefont {Xiang}},\ }\bibfield  {title}
  {\enquote {\bibinfo {title} {\textit{Second Renormalization of Tensor-Network
  States}},}\ }\href {\doibase 10.1103/PhysRevLett.103.160601} {\bibfield
  {journal} {\bibinfo  {journal} {Phys. Rev. Lett.}\ }\textbf {\bibinfo
  {volume} {103}},\ \bibinfo {pages} {160601} (\bibinfo {year}
  {2009})}\BibitemShut {NoStop}%
\end{thebibliography}%
\end{document}